\documentclass[12pt]{iopart}

\usepackage{graphicx}
\usepackage{dcolumn}
\usepackage{float}
\usepackage{iopams}

\newcommand{\cm}{cm$^{-1}$ }

\begin{document}


\title{\boldmath Doping dependent optical properties of
Bi$_{2}$Sr$_{2}$CaCu$_{2}$O$_{8+{\delta}}$ \unboldmath}

\author{J Hwang$^{1}$, T Timusk$^{1,2}$ and G D Gu$^{3}$}
\address{$^{1}$Department of Physics and Astronomy, McMaster University, Hamilton, ON L8S 4M1,
Canada\\ $^{2}$The Canadian Institute of Advanced Research, Toronto, Ontario M5G 1Z8,
Canada\\ $^{3}$Department of Physics, Brookhaven National Laboratory, Upton, NY
11973-5000, USA}

\ead{hwangjs@mcmaster.ca}

\date{\today}

\begin{abstract}
We report on the ab-plane reflectance  of the high temperature superconductor
Bi$_{2}$Sr$_{2}$CaCu$_{2}$O$_{8+{\delta}}$ (Bi-2212). Samples spanning the doping range
from under doped with $T_{c}$ = 67 K(UD), to optimally doped with $T_{c}$ =96 K (OPT), to
over doped with $T_{c}$ =60 K (OD) were measured from room temperature down to the
superconducting state.  The measured reflectance data were analyzed to extract the
optical conductivity and the real and imaginary parts of the free carrier optical
self-energy.  We get an estimate of the dc resistivity  from the low frequency
extrapolation of the optical conductivity and the superfluid density from the imaginary
part of the optical conductivity. The conductivity sum rule can be related to the changes
of the kinetic energy of the system. When this system becomes a superconductor, the
kinetic energy decreases in the underdoped samples and increases in overdoped ones. The
optical self-energy, obtained from the extended Drude model, is dominated by two channels
of interaction, a sharp mode and a broad background.  The amplitude of the mode is
strongly doping and temperature dependent whereas the background decreases weakly with
doping and is nearly temperature independent.
\end{abstract}

\pacs{74.25.Gz, 74.62.Dh, 74.72.Hs}

\maketitle


\section{Introduction}

Since the discovery of high-temperature superconductivity in the copper
oxides~\cite{bednorz86} the dynamical properties of the charge carriers in these
materials have been investigated with a variety of spectroscopic techniques. Within a
certain class of theories these properties provide a direct fingerprint to the the
mechanism of superconductivity. For example, if in analogy with the BCS superconductors,
superconductivity is driven by the exchange of bosons, then a study of the self-energy
spectrum of the superconducting current carriers would help us in identifying the
relevant bosons. Traditionally this has been done with tunneling spectroscopy where
numerical inversion techniques using Migdal-Eliashberg theory\cite{carbotte90} revealed
spectroscopic fingerprints in the self-energy that could be compared with the phonon
densities of states determined by inelastic neutron scattering and thereby unambiguously
identified as phonons~\cite{rowell-mcmillan69}. The self-energy spectra of the high
temperature superconductors have also been investigated with increasing  success with
advanced spectroscopic techniques~\cite{eschrig06} including angle resolved photoemission
(ARPES), tunneling, optics and magnetic neutron scattering, but so far at least no
consensus has been achieved in the interpretation of these spectra.

Contrasting with the traditional BCS mechanism of boson exchange there are alternative,
exotic models of the mechanism of high temperature superconductivity involving the
formation of new states at high temperature which would then Bose condense  to form a
superconductor as the temperature is lowered.  There have also been suggestions that
combine the two pictures where at low doping levels, close to the antiferromagnetic part
of the phase diagram, an exotic mechanism operates but with increasing hole concentration
a more conventional boson exchange takes over.  To throw some light on these issues we
have undertaken a study of the spectrum of excitations responsible for  the self-energy
of the carriers as a function of doping in the Bi-2212 system using the optical
conductivity as our experimental probe. The Bi-2212 system is has several advantages for
such an investigation.  First the system can be doped from the moderately underdoped
region with $T_c$=60 K through the optimally doped well into the overdoped region with
$T_c$=60 K. Furthermore, because the crystals of this material can be vacuum cleaved, they
have also been studied extensively with ARPES~\cite{Damascelli03} and
tunneling~\cite{mourachkine05}.

The extended Drude model that we use to find the real and imaginary parts of the
scattering rate and the self-energy assumes that there is only one channel of
conductivity and that any deviation from the Drude form is due to inelastic interactions.
The validity of this assumption can be tested to some extent by comparing the scattering
rate obtained from optics with the results from ARPES. When such comparisons are made the
resulting self-energy spectra generally agree to an accuracy of about 15
\%~\cite{kaminski00,hwang04}. One cannot expect any better agreement from the two methods
since they measure different things, ARPES gives us the quasiparticle lifetimes and self
energies for any given $k$ vector whereas the optical self-energy represents an average
over the Fermi surface of contributions to the current excited by the external field with
each point characterized by a velocity component parallel to the field and a life time of
this particular quasiparticle. It therefore comes as a surprise that the two methods give very
similar results not only for the absolute values of these quantities but also their
frequency and temperature dependencies, in particular if the ARPES results for
quasiparticles near the nodal point are compared with the optical conductivity. Away from
the nodes there are large discrepancies: at low temperatures ARPES quasiparticles do not
show the narrowing that the optical conductivity shows and there appears to be an inelastic
scattering background that is not seen in the optical conductivity.

Many optical studies have been done on the cuprate systems, especially
La$_{2-x}$Sr$_x$CuO$_4$ (LSCO) and YBa$_2$Cu$_3$O$_{6+x}$ (Y123)
for light polarized along the ab-plane as well as the less
conducting  c-axis direction.\cite{basov05} Much early optical
work has was also done on the important Bi-2212 material
~\cite{puchkov96c,puchkov96d,quijada99,liu99}.  Quijada {\it et
al.}~\cite{quijada99} give a complete set of references to the
earliest work. However recent advances in crystal growth with
image furnaces have yielded better quality crystals, suggesting a
need for new studies of this material.   Here we report on the
optical properties of the ab-plane of Bi-2212 over a broad doping
range from underdoped to overdoped at various temperatures above
and below $T_c$. We obtain various optical quantities from the
measured reflectance and discuss some important doping dependent
issues. In particular, we focus on the doping dependent superfluid
density and the doping dependence of the contributions of various
excitations to the optical self-energy.

The superfluid density is a fundamentally important quantity in superconductivity because
it is directly related to the ability of the superconducting wave function to resist
perturbations to its phase. The superfluid density can be obtained from the optical
spectrum in two ways, by analyzing the imaginary part of the conductivity as a function
of frequency, or by comparing the spectral weight loss in the low frequency region as the
superconducting condensate is formed, according to the Ferrell-Glover-Tinkham (FGT) sum
rule~\cite{ferrell58}. Any discrepancy in the superfluid density obtained by the two
methods can be related to the kinetic energy difference between normal and
superconducting states. This difference can in turn be related to proposed models of the
superconducting mechanism driven by kinetic energy~\cite{hirsch92,hirsch00,deutscher05}.
Recent interest has focussed on the {\it doping dependence} of the superfluid density and
the kinetic energy difference~\cite{hirsch00,deutscher05}.

The suggestion that spin fluctuations might provide the pairing bosons was made early by
the theoretical community~\cite{monthoux91,millis90}. Adding to the interest in magnetic
excitations was the absence of an isotope effect in optimally doped samples and the
discovery of the sharp magnetic resonance mode at 41 meV in Y123 in the magnetic
susceptibility spectrum by inelastic neutron
scattering~\cite{rossat-mignod91,mook93,fong95,mook98,dai99,fong99,he01,he02}. A sharp
excitation at this energy can also be seen by other experimental techniques such as
optical spectroscopy~\cite{hwang04,thomas88,puchkov96,carbotte99,abanov99,tu02},
ARPES~\cite{shen97,norman98,campuzano99,bogdanov00,johnson01,kaminski01}, or
tunneling~\cite{zasadzinski01,zasadzinski06}. This excitation has the interesting
property that its energy $\hbar\Omega$ is proportional to the superconducting transition
temperature according to  $\hbar\Omega\approx 5k_BT$ at all doping levels~\cite{he01} and
in all the cuprate families where large single crystals are available. The only exception are
the reduced $T_c$ systems such as LSCO where, while there is peak in the magnetic
suscpeptibilty, a sharp resonance does not develop at low temperature.  In the overdoped
samples the magnetic resonance appears only below $T_c$ but in underdoped samples it can
be seen well above $T_c$~\cite{dai99,stock03}. However even in these systems the
resonance grows rapidly in strength at the superconducting transition. This temperature
dependence of the mode can also be seen in ARPES spectra in the nodal
direction~\cite{johnson01,kordyuk05} In the view of all these connections to
superconductivity it is not surprising that the role of magnetic resonance mode in
superconductivity has been discussed widely~\cite{dai99,demler98,scalapino99,kee02}.

One possible connection of the magnetic resonance to superconductivity would be through
its contribution to the optical and quasiparticle self-energy, where it would have a role
similar to the role of phonons in BCS
superconductivity~\cite{hwang04,carbotte99,abanov99,norman98,johnson01,munzar99,abanov02,hwang05a}.
The problems with this scenario have been pointed out by several authors. In the
overdoped Y123 the resonance is not present in the normal state and cannot initiate the
transition at $T_c$.  In the highly overdoped state the signature of the resonance
vanishes from the optical conductivity at a doping level where $T_c$ is around 60
K~\cite{hwang04,puchkov96,ma06} in both Bi-2212 and Tl$_2$Ba$_2$CuO$_{6+\delta}$
(Tl-2201). A general weakening of the contribution of the magnetic resonance to the
carrier self-energy with doping has also been seen in ARPES and
tunneling~\cite{johnson01,zasadzinski06,kordyuk05}. It was suggested by Hwang {\it et
al}~\cite{hwang04} that the coupling of the resonance disappears above a critical doping
level of $p=0.24$.  Other experiments support the idea that there is a significant change
in the properties of the cuprates at this doping level.  Shibauchi {\it et
al.}~\cite{shibauchi01} observed that the pseudogap temperature $T^*$ in c-axis transport
merged to $T_c$ in Bi-2212 system near the critical doping, 0.24. Ozyuzer {\it et
al.}~\cite{ozyuzer02} have observed that there was no indication of a pseudogap near at
Fermi level in their tunnelling conductance spectrum of a very overdoped Bi-2212 with
$T_c$=56 K. Some ARPES studies showed that near the critical doping the topology of the
Fermi surface transformed from hole-like to electron-like in Pb-doped
Bi-2201~\cite{yoshida01,takeuchi01}, LSCO~\cite{ino02} and Bi-2212~\cite{kaminski05}.
Another ARPES study showed that a crossover two- to three-dimensional electronic
structure occurred near the critical doping with $T_c$= 22 K in
(Bi,Pb)$_2$(Sr,La)$_2$CuO$_{6+\delta}$ system~\cite{takeuchi05}, which has $T^{max}_c$=
35 K.

In addition to the sharp mode there is a continuous bosonic background spectrum that is
responsible for the strong linearly rising scattering rate that extends to very high
frequencies.  It has been discussed in the theoretical literature in the context of
models such as the marginal Fermi liquid (MFL)~\cite{varma89} or the interaction of the
charge carriers with a continuous spectrum of spin fluctuations the Millis-Monien-Pines
(MMP) spectrum~\cite{millis90}. This broad background exists at all temperatures and
doping levels and in all cuprate systems including LSCO system~\cite{hayden98}, where
there is no clear evidence for the presence of the magnetic resonance mode. This
ubiquitous feature in the cuprates may be also be responsible for the broad kink feature
in ARPES spectra near nodal region~\cite{lanzara01}. Other strongly correlated electron
systems have similar broad backgrounds in their bosonic spectra or the real part of the
optical self-energy, for an example, the sodium cobaltate system~\cite{hwang05b}.
Interestingly, the sharp mode and the broad background was captured by optical
spectroscopy in early days of the high temperature of superconductors~\cite{collins89}.

In the next section we describe our experimental  technique including sample preparation. In
the following section, we provide measured reflectance spectra and describe how we
extract the optical conductivity from the measured reflectance. In the following section,
we extract the superfluid density using two different methods and the kinetic energy
change going from the normal to the superconducting state. Next, we
introduce the optical self-energy using the extended Drude formalism and describe the
doping and temperature dependent properties of the self-energy. In the final section, we
we relate our experiments to several important recent issues and provide overall summary
of our work.

\section{Sample preparations and experiments}

The Bi-2212 single crystals used in the study were grown in an optical image furnace with
the traveling solvent floating-zone technique. To get the appropriately doped samples
from as-grown crystals we annealed under various oxygen-annealing
conditions~\cite{eisaki04}; which yielded good samples of  the underdoped (UD), optimally
doped (OPT), and overdoped (OD) phases. Unwanted c-axis longitudinal phonons can be
admixed in ab-plane optical data of other cuprates such as LSCO and
Y123~\cite{orenstein88,reedyk92}. We see no evidence of these in our samples.  We could obtain shiny optical quality ab-plane surfaces
Bi-2212 by cleaving the sample.

A commercial Fourier transform spectrometer, Bruker IFS 66v/S, was used to obtain the
reflectance data over a wide frequency range from 50 to 40,000 cm$^{-1}$. For the low
temperature measurements we used a continuous flow liquid Helium cryostat with an
automated temperature control and sample change system~\cite{hwang03} to improve the
reproducibility over a manual system. A polished stainless steel mirror was used as an
intermediate reference to correct for the instrument drift with time and temperature. An
in-situ evaporated gold (50 - 14,000 cm$^{-1}$) or aluminum (12,000 - 40,000 cm$^{-1}$)
film on the sample was used as the final reflectance reference~\cite{homes93}. The
reflectance of the gold films was in turn calibrated with a polished stainless steel
sample where we relied on the Drude theory and the dc resistivity as the ultimate
reference. An advantage of this method is that it corrects for geometrical effects of an
irregular surface. The in-situ gold evaporation technique gives accurate temperature
dependent data with an accuracy better than $\pm$ 0.05 \% at room temperature.

We measured reflectance of four overdoped ($T_{c}$ = 80 K, 82 K, 65 K, and 60 K), one
optimally doped ($T_{c}$ = 96 K), and one underdoped ($T_{c}$ = 69 K) Bi-2212 sample. We
estimated the hole doping levels of the samples using the empirical parabolic equation of
Presland {\it et al.}~\cite{presland91}: $p(T_{c}) = 0.16
\mp[{1}/{82.6}\,(1-{T_{c}}/{T_{c}^{max}})]^{1/2}$, where $T_{c}^{max}$ is the maximum
$T_{c}$ of the material. The determination of $T_{c}^{max}$ is not an easy
problem~\cite{eisaki04} as it depends on the  growth conditions and dopant levels. In the
absence of a better method, we use the generally accepted value of 91 K as the
$T_{c}^{max}$ for Bi-2212. We should mention here that our optimally-doped sample is
doped with a small amount of Y to yield a relatively well ordered system and
shows a surprisingly high $T_{c}=96$ K~\cite{eisaki04} and we assigned $p=0.16$ as its
hole doping level. The disadvantage of the parabolic approximation is that it does not
uniquely determine the doping level of the sample since there are two independent $p$
values for each value of $T_{c}$. To avoid the ambiguity in the determination of the
doping levels we used the slope of the infrared reflectance as an additional test of the
doping level~\cite{hwang04a}. Table~\ref{plasma} shows the estimated doping levels of our
eight Bi-2212 samples.

\section{Reflectance and optical conductivity}

In this section we provide the measured reflectance and the
optical conductivity derived from the reflectance by using
Kramers-Kronig (KK)
transformations~\cite{quijada99,wooten72,santander02,kuzmenko05a}.
To use this method we have to extrapolate the measured reflectance
to zero frequency on the low frequency side of the measured range
and to infinite frequency on the high frequency side. We used the
following extrapolations. For $\omega \rightarrow 0$, the
reflectance was extrapolated by assuming a Hagen-Rubens frequency
dependence, $1-R(\omega)\propto \omega^{1/2}$ for normal states
and $1-R(\omega)\propto \omega^{4}$ for the superconducting
states. For $\omega \rightarrow \infty$, the reflectance has been
extended by using a literature data~\cite{tarasaki90} and
free-electron behavior ($R\propto \omega^{-4}$).
%
%
\begin{figure}[t]
  \centering
  \vspace*{-1.5cm}
  \centerline{\includegraphics[width=4.0in]{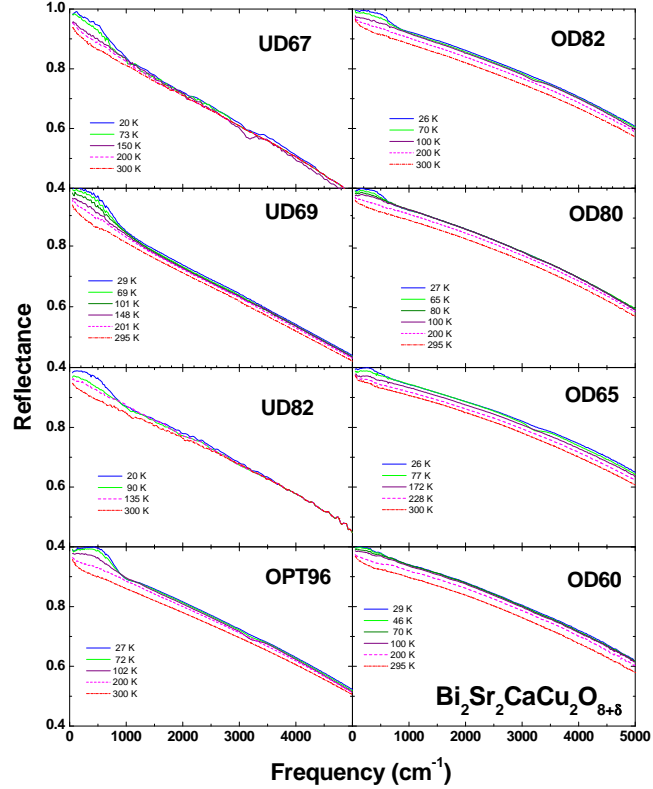}}
  \vspace*{-1.0cm}
  \caption{ab-plane reflectance of Bi-2212 at eight different doping levels.
The overall reflectance increases as the hole doping level increases with charge carrier
density introduced by doping. Here and in the following figures we use a brief notation
to describe doping status of our samples, for example UD67 stands for the underdoped
sample with $T_c$=67 K, OD for overdoped and OP for optimal doping. We note that a dip
feature near 3300 \cm of the low temperature data of some of the samples an absorption is
a band from ice on the sample. This feature does not affect low frequency data below the
absorption frequency, 3300 cm$^{-1}$. However it affects the high frequency data above
3300 cm$^{-1}$.} \label{fig1}
\end{figure}
%
%
\begin{figure}[t]
  \centering
  \vspace*{-1.5cm}
  \centerline{\includegraphics[width=4.0in]{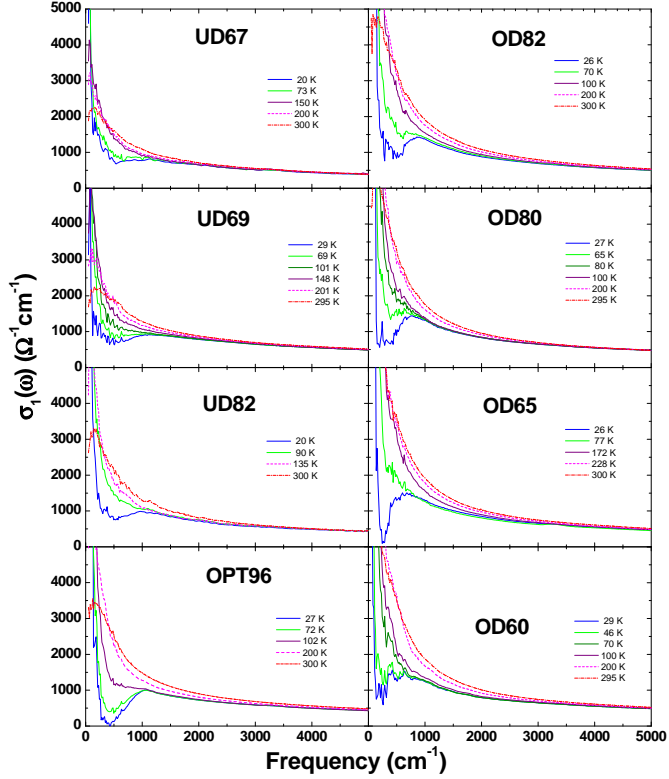}}
  \vspace*{-1.0cm}
  \caption{The ab-plane optical conductivity of Bi-2212 of eight
Bi-2212 samples at various temperatures above and below $T_c$.}
\label{fig2}
\end{figure}

To obtain further optical constants from the measured reflectance ($R(\omega)$) and the
corresponding calculated phase ($\phi(\omega)$) from KK analysis, we used the Fresnel's
equation for normal incidence:
\begin{equation}
\frac{1-n(\omega)-ik(\omega)}{1+n(\omega)+ik(\omega)}
=\sqrt{R(\omega)}e^{i\phi(\omega)}
\end{equation}
or
\begin{eqnarray}
n(\omega)&=&{{1-R(\omega)} \over  {1+R(\omega)
-2\sqrt{R(\omega)}\cos{\phi(\omega)}}}\\
k(\omega)&=&{{-2\sqrt{R(\omega)}\sin{\phi(\omega)}} \over
{1+R(\omega)-2\sqrt{R(\omega)}\cos{\phi(\omega)}}}
\end{eqnarray}
where $n(\omega)$ and $k(\omega)$ are the index of refraction and the extinction
coefficient, respectively.

In principle, we are able to obtain any other optical quantities by using the various
relationships between the optical quantities~\cite{wooten72}.
\begin{eqnarray}
\epsilon_1(\omega)&=&n^2(\omega)-k^2(\omega)\\
\epsilon_2(\omega)&=&2 n(\omega)k(\omega)
\end{eqnarray}
and
\begin{equation}
\sigma(\omega)=-i\frac{\omega}{4
\pi}[(\epsilon_1(\omega)-\epsilon_{H})+i\epsilon_2(\omega)]
\end{equation}
where $\epsilon_1(\omega)$ and $\epsilon_2(\omega)$ are the real
and imaginary parts of the optical dielectric constant,
respectively, and $\sigma(\omega)\equiv
\sigma_1(\omega)+i\sigma_2(\omega)$ is the complex optical
conductivity.

The quantity $\epsilon_{H}$ is the contribution to the dielectric constant from all the
high frequency spectral weight but excluding the low frequency free carrier part or
intraband spectral weight. It is often difficult to determine where the dividing line
between the free carriers and the interband absorption lies.  We take $\epsilon_H$ at a high
frequency ($\sim 2 eV$) for each doping level (note that there is doping dependence in
$\epsilon_H$). To get $\sigma_2$ we used $\epsilon_H$ values shown in
table~\ref{plasma}. Recently, the accuracy of the various methods of finding  $\epsilon_H$ has becomes an important issue because the optical scattering rate is very sensitive to the value of $\epsilon_H$, especially at
high frequency (for more discussions see the following section and figure~\ref{fig6} and
the related text). %
%
\begin{figure}[t]
  \centering
  \vspace*{-1.5cm}
  \centerline{\includegraphics[width=4.0in]{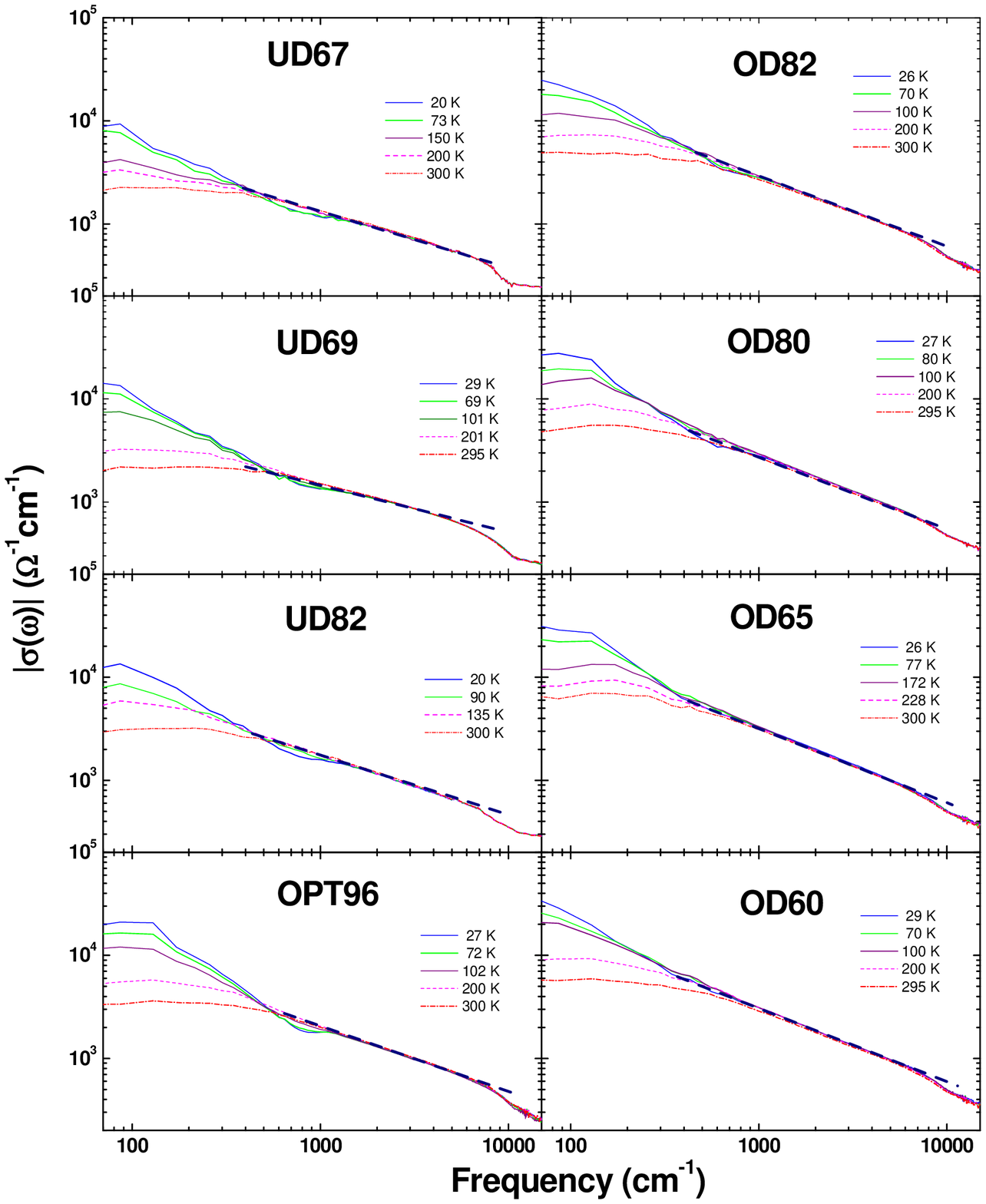}}
  \vspace*{-1.0cm}
  \caption{Amplitude of the optical conductivity,
$|\sigma(\omega)|\equiv
\sqrt{\sigma_{1}^{2}(\omega)+\sigma_{2}^{2}(\omega)}$. The dashed
lines are least square fits of the temperature independent
region.} \label{fig3}
\end{figure}

%
%
\begin{figure}[t]
  \centering
  \vspace*{-1.5cm}
  \centerline{\includegraphics[width=4.0in]{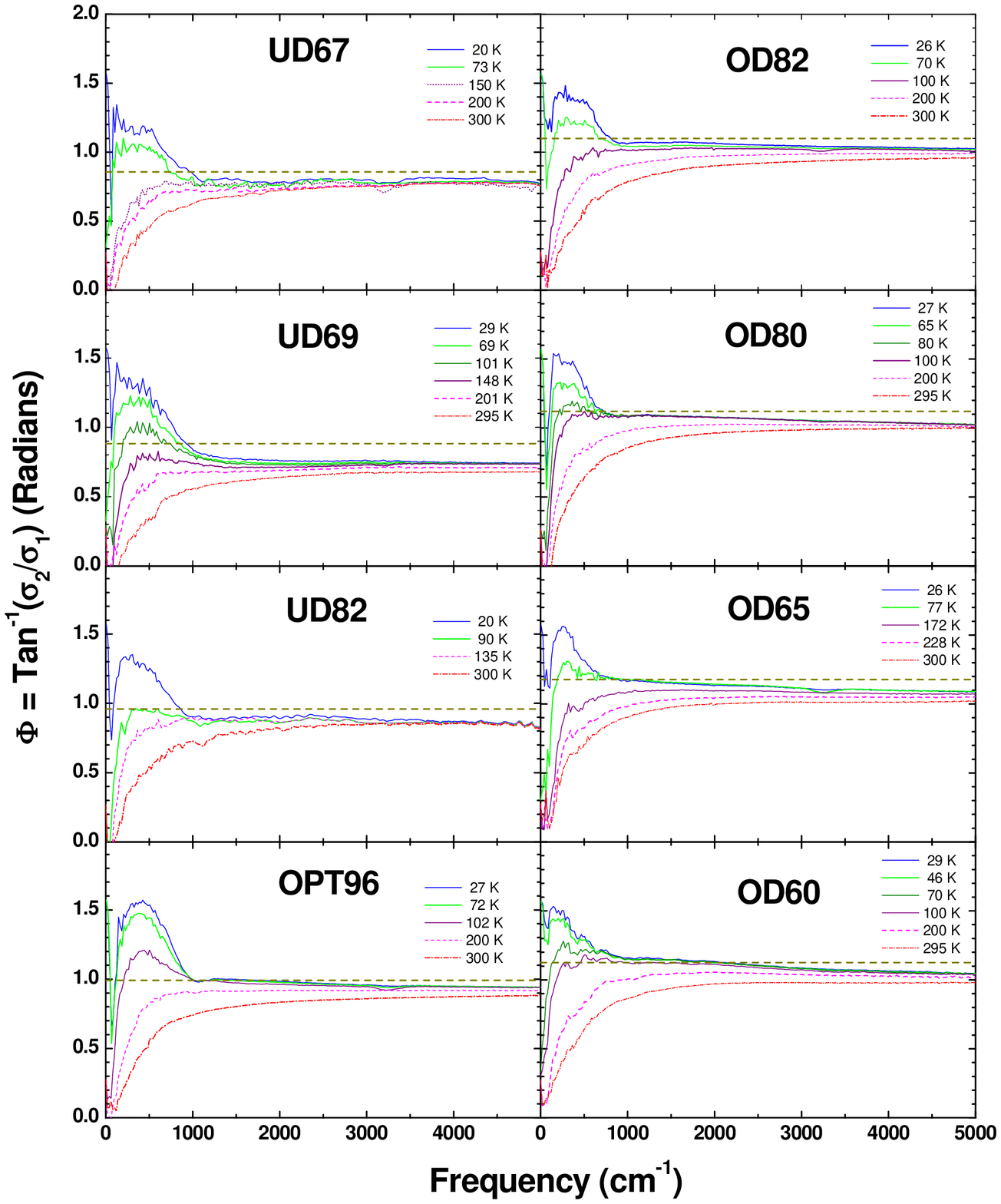}}
  \vspace*{-1.0cm}
  \caption{Phase of the optical conductivity, $\Phi(\omega)
\equiv$tan$^{-1}({\sigma_{2}(\omega)/\sigma_{1}(\omega)})$~\cite{vanderMarel04}.
The dashed horizontal lines are obtained from the fitted straight
lines in Fig.~\ref{fig3}.} \label{fig4}
\end{figure}

Figure~\ref{fig1} displays the measured ab-plane reflectance of Bi-2212 at eight different
doping levels and at various temperatures. As the hole doping level increases, some
interesting doping dependent features show up; the overall reflectance level increases
and the overall shape becomes more curved. For each doping level, as the temperature decreases
the overall reflectance increases because of the reduced scattering and, more interestingly, a
shoulder appears between 500 \cm and 1000 \cm below an onset
temperature, which is also doping dependent. We will analyze the shoulder feature in
detail and discuss its doping dependent properties in the section  "Extended Drude
model and optical self-energy".  At room temperature the reflectance below 2000 \cm is
approximately linear in frequency. This linear variation, the so called marginal Fermi
liquid behavior~\cite{varma89} in Bi-2212 at 300 K has been analysed in detail by Hwang
{\it et al.}~\cite{hwang04a} below 2000 \cm over a wide doping range.

Figure~\ref{fig2} displays the real part of the optical conductivity of Bi-2212. We
observe strong temperature dependence only in low frequency region, below 3000 cm$^{-1}$.
The spectral weight or the number of effective charge carriers per Cu atom on CuO$_2$
plane is defined as follows:
\begin{equation}
N_{eff}(\omega)=\frac{2 m V_{Cu}}{\pi
e^2}\int_{0^{+}}^{\omega}\sigma_{1}(\omega')d\omega'
\end{equation}
where $m$ is the free electron mass, $V_{Cu}$ is the volume per Cu atom in the sample,
$e$ is the charge of an electron, and $\sigma_{1}$ is the real part of the optical
conductivity. The spectral weight is proportional to the area under $\sigma_1(\omega)$
curve. As the hole doping increases the overall conductivity increases. For each doping
level,  as the temperature decreases, the spectral weight at low frequencies increases
as a result of the narrowing the Drude-like absorption band near zero frequency. This shift of
spectral weight continues down to the superconducting transition temperature, $T_c$.
Below or near $T_c$ a strong depression of spectral weight below 1000 \cm sets in and
grows with decreasing temperature. This feature is related to the step-like feature in
the optical scattering rate. (for more detailed discussion see Sec. "Extended Drude model
and optical self-energy") The missing area between the normal and the superconducting curves can
be a measure of the superfluid density (more detailed discussion see the section "FGT Sum
rule, superfluid density, and kinetic energy"). Even in the superconducting state we have a
sizable amount of residual spectral weight near zero frequency, which is absent in the
conventional superconductors~\cite{mattis58,tinkham75,leplae83,zimmermann91}.

%
%
\begin{figure}[t]
  \centering
  \vspace*{-2.5cm}
  \centerline{\includegraphics[width=4.0in]{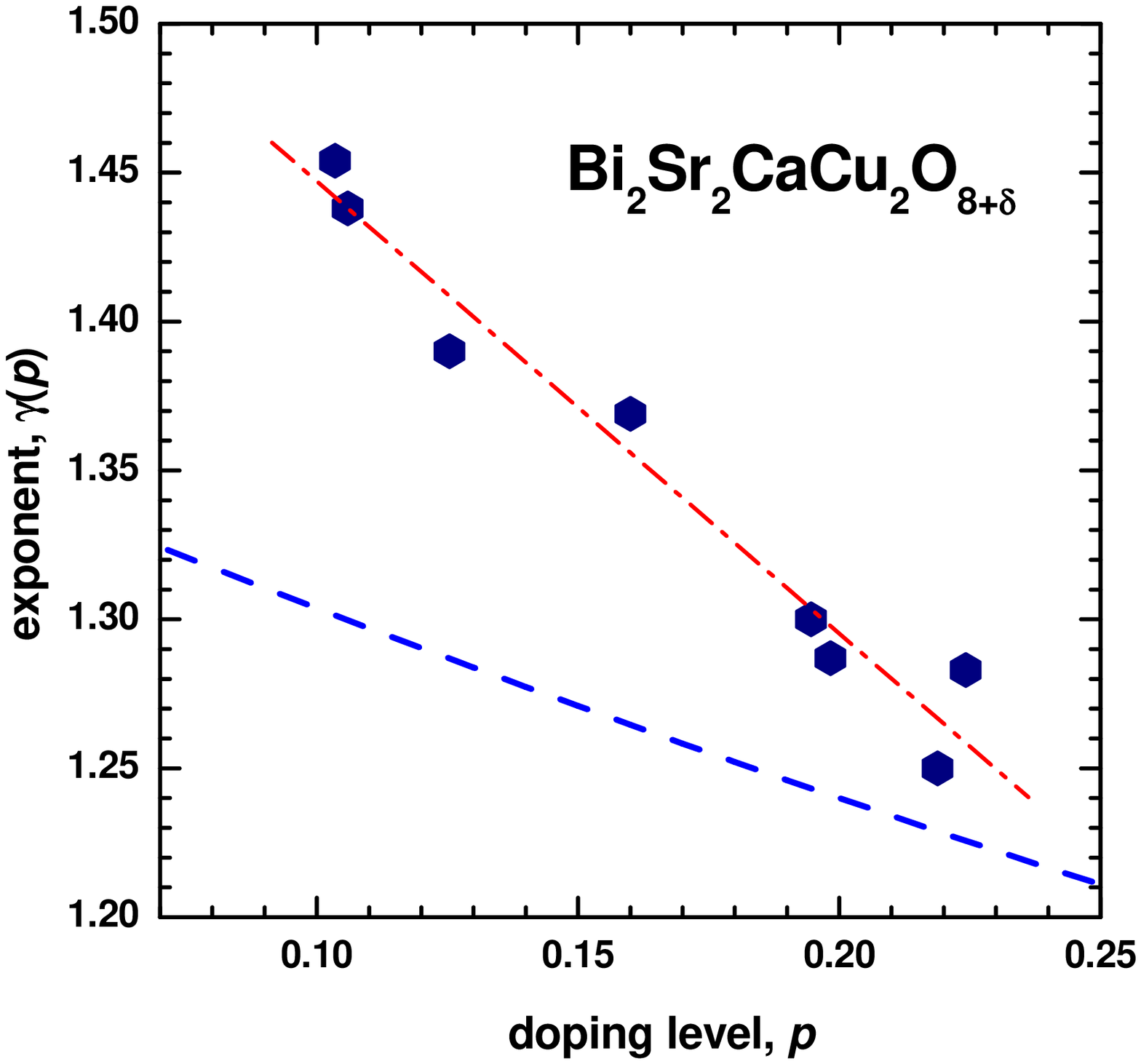}}
  \vspace*{-2.5cm}
  \caption{Doping dependent exponent, $\gamma(p)$. See the text
for a detail description of the quantity $\gamma(p)$. The dashed
line is a  theoretical estimate of the variation of $\gamma(p)$
with doping due to Anderson, ref.~\cite{anderson05}} \label{fig5}
\end{figure}

%
%
\begin{figure}[t]
  \centering
  \vspace*{-1.5cm}
  \centerline{\includegraphics[width=4.0in]{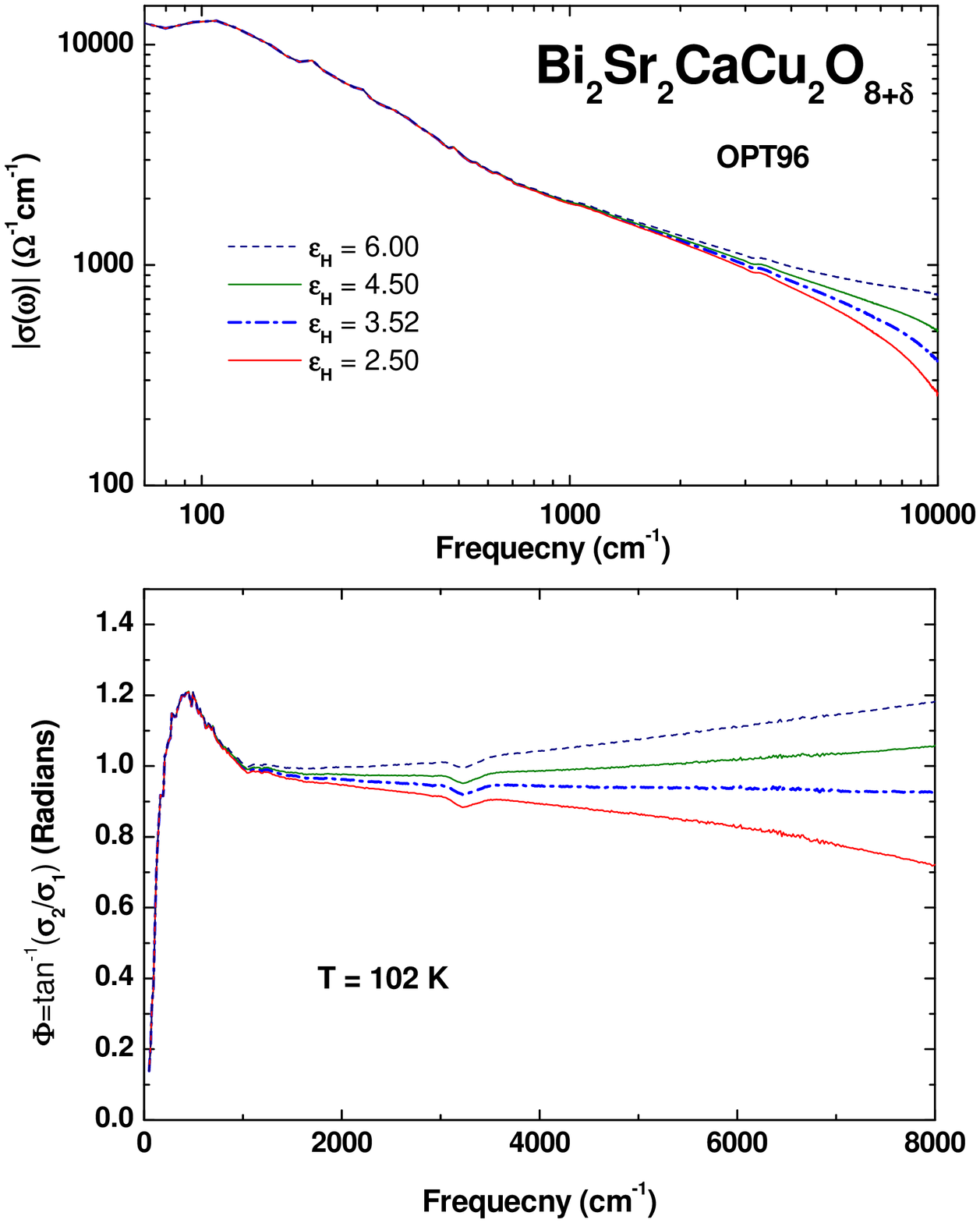}}
  \vspace*{-1.0cm}
  \caption{The effect of choosing different values of $\epsilon_{H}$
on the amplitude a) and the phase b) of the optical conductivity.
Note that  $\sigma_1$ is independent of  $\epsilon_{H}$.}
\label{fig6}
\end{figure}

The complex optical conductivity can be described by an amplitude and a phase, i.e.
$\sigma(\omega)=|\sigma(\omega)|e^{i\Phi(\omega)}$, where $|\sigma(\omega)| =
\sqrt{\sigma_{1}^{2}(\omega)+\sigma_{2}^{2}(\omega)}$, and $\Phi(\omega)=$
tan$^{-1}(\sigma_2({\omega})/\sigma_1({\omega}))$. Anderson~\cite{anderson97} suggested
simple power law behavior for the complex optical conductivity in his Luttinger Liquid
model. Recently van der Marel {\it et al.}~\cite{vanderMarel04} revisited the Anderson's
model describing the optical conductivity as $\sigma(\omega)= C (-i\omega)^{\gamma-2}$,
where $C$ is a frequency independent constant. In this description we have
$|\sigma(\omega)|=C \omega^{\gamma-2}$ and $\Phi(\omega)=\pi/2(2-\gamma)$, which is
frequency independent. Note that this formalism can describe only the temperature independent
parts of the conductivity spectra. In figure~\ref{fig3} we plot the amplitude of the
optical conductivity of our eight Bi-2212 samples at various temperatures on a log-log
scale. The dashed lines are linear least squares fits of the amplitude between 500 \cm
and 5000 \cm in the normal state where the quantity $|\sigma(\omega)|$ shows negligible temperature dependence.
The absolute value of the slope ($\gamma-2$) of the fitted straight line increases as the
doping level increases. From the slope we can obtain the doping dependence of
$\gamma(p)$, which decreases monotonically from 1.45 to 1.25 as doping increases (see
figure~\ref{fig5}).

In figure~\ref{fig4} we show the phase, $\Phi(\omega)$, of the optical conductivity for
all our samples. We note that the phase is almost frequency independent over a wide
spectral range above 1000 \cm at all measured temperatures. The dashed horizontal line is
the corresponding phase of the straight line fit of the amplitude (see
figure~\ref{fig3}). We observe that the van der Marel behavior holds over a wide spectral
range, roughly between 500 \cm and near 5000 cm$^{-1}$, at various temperatures, and over a
wide rage of doping. We also note that at low frequencies, below 500 cm$^{-1}$, the
optical conductivity shows a strong temperature dependence and deviates markedly from a
constant. In figure~\ref{fig5} we plot the doping dependent exponent, $\gamma(p)$, which
is extracted from the linear fits in figure~\ref{fig3}. The exponent decreases
monotonically as the doping increases. The dot-dashed line is a list squares fit of a
straight line to the data.  The lower dashed line is a predicted variation of this
quantity with doping from a recent paper by Anderson\cite{anderson05}.

In figure~\ref{fig6} we demonstrate how the value $\epsilon_{H}$ affects the amplitude
and the phase of the optical conductivity. Note that $\epsilon_{H}$ affects  only the
imaginary part of the optical conductivity, $\sigma_2$. To see the effect we chose the
data for optimally doped Bi-2212 at $T$=102 K and calculate the complex optical
conductivity for four different values of $\epsilon_{H}$ between 2.50 and 6.00. It is
clear from  the figure that there is no significant effect below 1000 cm$^{-1}$ but the
effect builds up rapidly as the frequency increases. In the upper panel we show
calculated amplitudes on a log-log scale. As $\epsilon_{H}$ increases the absolute value
of the slope in the amplitude between 500 and 5000 \cm decreases. When
$\epsilon_{H}$=3.52, the value we used in our analysis, the amplitude is the straightest
over a wide range of frequencies. In the lower panel we show corresponding phases. As
$\epsilon_{H}$ increases the slope in the phase changes from negative to positive. When
$\epsilon_{H}$=3.52 the phase is almost frequency independent, i.e. the slope is zero.
However to reach a definite conclusion from this type of analysis~\cite{vanderMarel04}
one needs a model-independent way of estimating  $\epsilon_{H}$.

%
%
\begin{table}
\caption{\label{plasma} Doping levels ($p$), the background
dielectric constant ($\epsilon_H$), and the plasma frequency
($\omega_p$) of our eight Bi-2212 samples.}
\begin{indented}
  \item[]\begin{tabular}{|c|c|c|c|c|c|}\hline
 Samples&$T_{c}(K)$&Doping level,$ p $&$\epsilon_H$&$\omega_p$ (cm$^{-1}$)\\\hline
     UD67&67&0.103&3.14&15 050\\\hline
     UD69&69&0.106&3.15&15 150\\\hline
     UD82&82&0.125&3.21&16 300\\\hline
     OPT96&96&0.160&3.52&17 000\\\hline
     OD82&82&0.195&4.14&18 800\\\hline
     OD80&80&0.198&4.15&18 700\\\hline
     OD65&65&0.219&4.25&19 200\\\hline
     OD60&60&0.224&4.19&18 900\\\hline
   \end{tabular}
\end{indented}
\end{table}

%
%
\begin{figure}[t]
  \centering
  \vspace*{-2.5cm}
  \centerline{\includegraphics[width=4.0in]{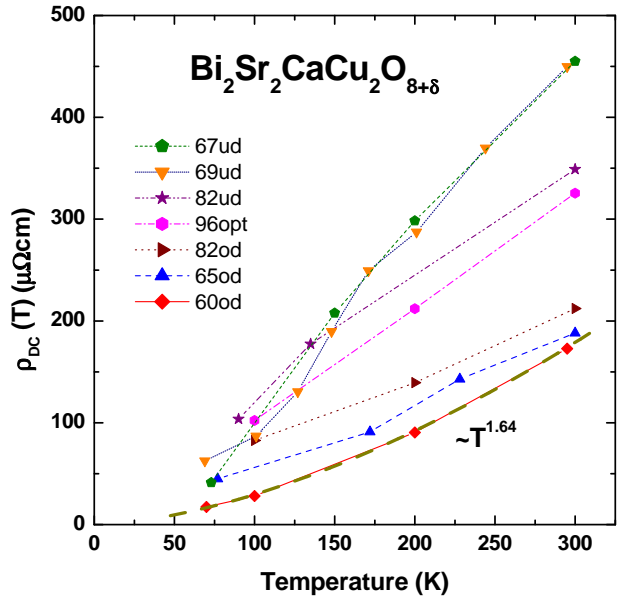}}
  \vspace*{-2.5cm}
  \caption{dc resistivity, $\rho_{dc}(T)$, extracted from low
frequency extrapolations of normal state optical conductivities.
The dashed curve through data points of our most overdoped sample
is a fitted curve, $\rho_{DC}(T)=0.0155\: T^{1.64}$. See the text
for a detailed description of the extrapolation.} \label{fig7}
\end{figure}


In figure~\ref{fig7} we show the temperature dependence of the dc
resistivity of our eight Bi-2212 samples obtained from
extrapolations to zero frequency of the normal state optical
conductivities. These optically determined resistivities show the
same features and trends that are seen with four-probe dc
measurements. Starting with the optimally doped sample OPT96 we
see the familiar linear variation of resistivity with temperature
with a zero temperature intercept that is very close to the zero
of the temperature axis. From the beginning of the study of the
cuprates this behavior has been taken as evidence of an exotic
transport mechanism. Interaction with a bosonic mode would also be
linear with temperature in this range but would yield an intercept
at approximately $\hbar\omega/4$ where $\omega$ is the frequency
of the mode.  For phonon modes involving oxygens this intercept
would be at 100 K. Moving to underdoped samples we do find a
higher intercept associated with the temperature dependence below
about 150 K.  Above this temperature the temperature dependence is
similar to that of the optimally doped sample with a lower
intercept on the temperature axis. The overdoped samples on the
other hand show a more metallic temperature dependence with an
upward curvature with the power law  index of 1.64 for our most
overdoped sample smaller than the $T^2$ expected for a Fermi
liquid. These observations are consistent all previous systematic
measurements of the dc resistivity~\cite{takenaka94,sutherland04}.
The message from the underdoped samples is that any bosonic
spectral function causing this scattering has a temperature
dependent amplitude with rapid changes occurring in the 150 K
temperature range.  This is in agreement with our study of another
underdoped system Ortho II YBCO~cite~\cite{hwang05a}.

\section{FGT sum rule, superfluid density and kinetic energy}

Of the various techniques used to determine the superfluid density optics has the
advantage that it gives the absolute value and in an anisotropic system, such as the high
temperature cuprates, all the orthogonal components if an oriented crystal is used.
However there are problems that make an accurate determination of the absolute magnitude
difficult. The first of these is a need to determine accurately the reflectance in the low
frequency region where the it approaches unity and where measurements on small
crystals are difficult as the wavelength of the radiation used approaches the size of the
samples.

Another source of uncertainty is the presence of a residual metallic conductivity in the
superconducting state seen in many samples.  This effect can be recognized by a downturn
in reflectance at low frequency by as much as one to two percent. The resulting
conductivity shows a peak removed from zero frequency characteristic of a disordered
metallic material.~\cite{basov93a} This low frequency suppression of reflectance is
directly related to a sizable low frequency spectral weight in the optical conductivity
and might cause some uncertainty in the estimated superfluid density. Here we try to minimize this uncertainty by estimating the
super fluid density with the same criteria for all our Bi-2212 samples using the same experimental
instrument and technique, the same method of analysis, and the same low frequency cutoff
of reflectance data, 50 cm$^{-1}$. With eight different data sets of the optical
conductivity of Bi-2212, we obtain doping dependent superfluid density from the highly
underdoped region to the overdoped region.

%
%
\begin{figure}[t]
  \centering
  \vspace*{-1.5cm}
  \centerline{\includegraphics[width=4.0in]{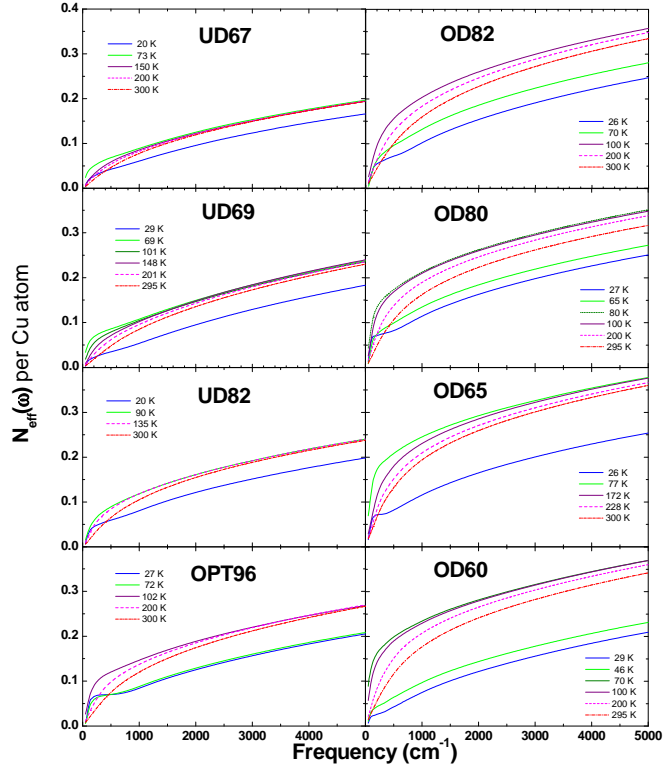}}
  \vspace*{-1.0cm}
  \caption{Frequency dependent spectral weight expressed as the number of effective charge
carriers per Cu atom of Bi-2212, as is defined in Eg. 7.}
  \label{fig8}
\end{figure}

Figure~\ref{fig8} shows the frequency dependent accumulated spectral weight or the
effective number of charge carriers per Cu atom of eight Bi-2212 samples at various
temperatures below and above $T_c$. Overall, the spectral weight increases as we add
carriers by doping as we expect. In the  normal state, as the doping level increases the
temperature dependent region in the spectral weight extends to higher frequency. As the
temperature is lowered spectral weight is shifted to low frequency because of the narrowing of the
Drude band.

Above 2000 \cm the highest spectral weight curves are almost parallel to the curve in the
superconducting state. The difference between the two parallel spectral weight curves
gives us a rough estimate of the number of superfluid charge carriers per Cu atom, the superfluid density. This difference
increases as the doping level increases; the higher the doping level, the higher is the
superfluid density (see figure~\ref{fig11}). The flat region spectral weight in the
superconducting state is related to the dip below 1000 \cm in the optical conductivity.
We analyze spectral weight variations in detail using the optical sum
rule~\cite{wooten72,tinkham75} with or without including the temperature dependence in
the spectral weight.

We describe next the method we use to obtain the superfluid density from the optical
conductivity in the superconducting state. In the superconducting state we have two
separate contributions to the real part of the optical conductivity: a delta function at
the origin from the superconducting condensate and a regular non-superconducting part.
\begin{equation}
\sigma_{1}(\omega)=\sigma_{1s}\delta(0)+\sigma_{1}^{'}(\omega)
\label{e1}
\end{equation}
where $\sigma_{1s}$ is a contribution from the superfluid charge carriers and
$\sigma_{1}^{'}(\omega)$ is a regular non-superconducting part of the optical
conductivity. From the optical sum rule, we can describe the superfluid plasma frequency,
$\omega_{ps}$ in terms of $\sigma_{1s}$; $\sigma_{1s}={\omega_{ps}^{2}}/{8}$. Note that
dimension of $\sigma_{1s}$ is frequency squared. The Kramers-Kronig (KK) transformation
of the real part gives us the imaginary part, which also consist of two terms, as
follows:
\begin{eqnarray}
\sigma_{2}(\omega)&\equiv&\sigma_{2s}(\omega)+\sigma_{2}^{'}(\omega)\label{e2} \\
&=&\frac{2\sigma_{1s}}{\pi \omega}+\sigma_{2}^{'}(\omega) \label{e3}
\end{eqnarray}
where $\sigma_{2s}(\omega)$ is the imaginary part of the condensate conductivity and
$\sigma_{2}^{'}(\omega)$ the imaginary part of the regular non superconducting part.  The
latter makes a KK pair with  $\sigma_{1}^{'}(\omega)$.
%
%
\begin{figure}[t]
  \centering
  \vspace*{-2.5cm}
  \centerline{\includegraphics[width=3.8in]{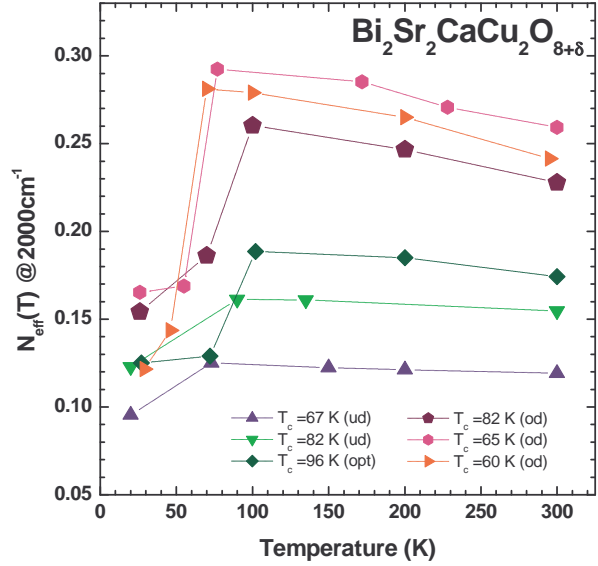}}
  \vspace*{-2.5cm}
  \caption{Temperature dependent spectral weight below2000
cm$^{-1}$ for six representative Bi-2212 samples. At  this cutoff
frequency the spectral weight increases as the temperature is
lowered to drop rapidly at $T_c$ as the superconducting condensate
forms. There is a monotonic increase of spectral weight with
doping.} \label{fig9}
\end{figure}

Using KK transformations we can calculate $\sigma_{2}^{'}$ from $\sigma_{1}^{'}$, which
is simply the optical conductivity in the superconducting state because $\sigma_{1s}$ is
a delta function at zero frequency and is not seen at finite frequencies. We  have
already extracted the total $\sigma_{2}$ from the measured reflectance and a given
$\epsilon_{H}$ by using KK analysis as described previously. So we can obtain
$\sigma_{2s}$ by subtracting $\sigma_{2}^{'}$ from $\sigma_{2}$. Finally, we can
calculate the superfluid plasma frequency, which is closely related to the superfluid
density, $N_{s}$, by the following equation:
\begin{equation}\sigma_{2s}(\omega) \cdot \omega= \frac{2\sigma_{1s}}{\pi} =
\frac{\omega_{ps}^{2}}{4 \pi} =\frac{N_{s}e^{2}}{m V_{Cu}}
\label{e7}
\end{equation}
where $V_{Cu}$ is the volume per Cu atom, $e$ is the charge of an electron, and $m$ is
the mass of a free electron.

After some units conversions, we get the following practical formula for the superfluid
density, $N_{s}$.
\begin{equation}
N_{s}\cong6.29\times10^{-10} V_{Cu}\cdot \sigma_{2s}(\omega)\cdot
\omega \label{e8}
\end{equation}
where $V_{Cu}$ is in $\AA^{3}$, $\sigma_{2s}$ in
$\Omega^{-1}$cm$^{-1}$, and $\omega$ in cm$^{-1}$. For Bi-2212
$V_{Cu}$ is 112.6 $\AA^{3}$, which does not change much with
doping.

Another independent method that yields the superfluid density from
the optical conductivity is directly related to the 'missing
spectral weight' in the optical conductivity when charge carries
are paired and condense to form the superfluid. There is a
fundamental difference between this and the previous method. For
this method we need to have two optical conductivities (one in
normal state and the other in superconducting state) while only
one optical conductivity in the superconducting state is needed
for the application of the previous method. We should note that
the temperatures of the normal and superconducting states are not
the same. The temperature difference may cause an unwanted error
in the superfluid density extracted using this methods because the
normal state spectral weight shows a temperature dependence (see
figure~\ref{fig9})~\cite{hwang05a,ortolani05} which will extend
below the superfluid transition temperature. This method
originates in the Ferrel-Glover-Tinkham (FGT) sum
rule~\cite{ferrell58} which states that all the spectral weight
lost at finate frequencies is transferred to the superconducting
condensate delta function at zero frequency. In the formalism we
can describe the superfluid plasma frequency in convenient units
as follows:
\begin{eqnarray}
\omega_{ps,sum}^{2}(\omega)&=&\frac{120}{\pi}\int_{0^+}^{\omega}
[\sigma_{1,n}(\omega')-\sigma_{1,s}(\omega')]d\omega'
 \label{e9} \\ \nonumber
N_{s,sum}(\omega)&\cong& 4.26 \times 10^{-10}\cdot V_{Cu}
\int_{0}^{\omega}(\sigma_{1,n}-\sigma_{1,s})d\omega'
\end{eqnarray}
where $\sigma_{1,n}$ and $\sigma_{1,s}$ are the real part of the
optical conductivity of normal and superconducting states,
respectively. $V_{Cu}$ is in $\AA^{3}$ and $\sigma_{1}$'s are in
$\Omega^{-1}$cm$^{-1}$. All frequencies are in cm$^{-1}$ including
$\omega_{ps,sum}$.

In conventional superconductors, where the pairing and the condensation are driven by
potential energy, the FGT sum rule holds exactly. However, as Hirsch has proposed, the FGT
sum rule can be violated in unconventional superconductors including the
cuprates~\cite{hirsch92} and a modified FGT sum rule can be introduced  with an extra
kinetic energy terms as follows~\cite{hirsch00,norman03}:
\begin{eqnarray} \nonumber
\omega_{ps}^{2}(\omega)\!\!&=&\!\!\frac{120}{\pi}\!\int_{0^+}^{\omega}
\!\!\Delta\sigma_{1}(\omega')d\omega'+ \frac{e^{2}a
b}{\pi\hbar^{2}c^{2}V_{Cu}}\Delta E_{Kin}(\omega)
 \label{e11} \\ \nonumber &\mbox{or}& \\
\Delta E_{Kin}(\omega)\!\!&=&\!\!\frac{\hbar^{2}}{m a
b}[N_{s}-N_{s,sum}(\omega)] \\ \nonumber &\mbox{or}& \\
\Delta E_{Kin}(\omega)\!\!&\cong& \!\!0.261\times
[N_{s}-N_{s,sum}(\omega)] \;\; (\mbox{in eV}) \label{e12}
\end{eqnarray}
where $\Delta\sigma_{1}(\omega')=\sigma_{1,n}(\omega')-\sigma_{1,s}(\omega')$, $a$ and
$b$ are the lattice constants of the ab-plane, $c$ is the speed of light, and $\Delta
E_{Kin}(\omega)$ is the kinetic energy change when the system becomes a superconductor,
i.e. $\Delta E_{Kin}(\omega)\equiv E_{Kin}^{s}(\omega)-E_{Kin}^{n}(\omega)$, where
$E_{Kin}^{s}(\omega)$ and $E_{Kin}^{n}(\omega)$ are kinetic energy of the superconducting and the
normal states, respectively. Note that Eq. 15 holds only for Bi-2212 systems.

%
%
\begin{figure}[t]
  \centering
  \vspace*{-2.0cm}
  \centerline{\includegraphics[width=4.0in]{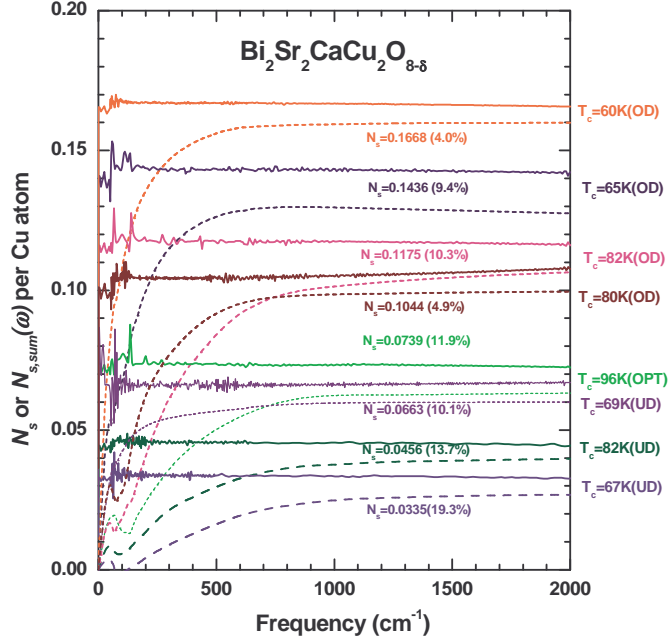}}
  \vspace*{-2.0cm}
  \caption{Superfluid densities obtained by two different
methods using $\sigma_2(\omega)$ (Eq.~\ref{e8}) (solid curves) and using the FGT
sum rule (Eq.~\ref{e9}) (dashed curves). See table~\ref{sample} for detailed
values of $N_{s}$.} \label{fig10}
\end{figure}

%
%
\begin{figure}[t]
  \centering
  \vspace*{-1.5cm}
  \centerline{\includegraphics[width=4.0in]{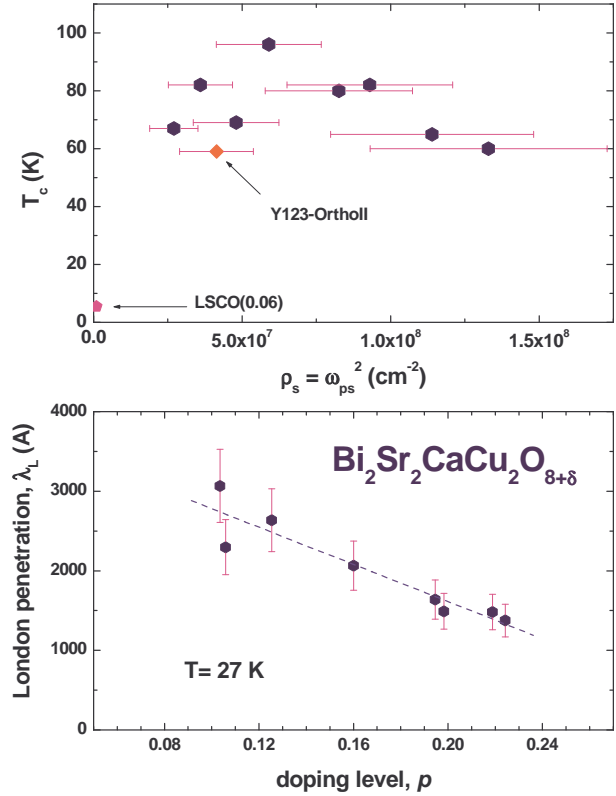}}
  \vspace*{-1.0cm}
  \caption{In the upper panel we display $T_c$ versus $\rho_{s}\equiv
\omega_{ps}^2$. We added two data point from underdoped LSCO with $T_c$= 5.5 K and
underdoped Y123 in the orthoII phase with $T_c$= 59 K. In the lower panel we display
doping dependent London penetration depth, $\lambda_L(p)$.} \label{fig11}
\end{figure}

%
%
\begin{table}
\caption{\label{sample} Superfluid densities of our eight Bi-2212 samples. $N_{s}$ are
the extracted superfluid densities using Eq.~\ref{e8} and $N_{s,sum}$ are the extracted
superfluid densities using FGT sum rule, Eq.~\ref{e9}, integrated up to 2000 cm$^{-1}$.
$N_{s,diff}(\%)$ are the percentage differences between superfluid densities determined
by the two methods (see in the text for a more detailed description).}
\begin{indented}
 \item[]\begin{tabular}{|c|c|c|c|c|c|}\hline
 Samples&$T_{c}(K)$&$p$&$N_{s}$&$N_{s,sum}$&$N_{s,diff}(\%)$\\\hline
     UD67&67&0.103&0.0335&0.0270&19.3\\\hline
     UD69&69&0.106&0.0663&0.0596&10.1\\\hline
     UD82&82&0.125&0.0456&0.0394&13.7\\\hline
     OPT96&96&0.160&0.0739&0.0651&11.9\\\hline
     OD82&82&0.195&0.1044&0.0993&4.9\\\hline
     OD80&80&0.198&0.1175&0.1054&10.3\\\hline
     OD65&65&0.219&0.1436&0.1301&9.4\\\hline
     OD60&60&0.224&0.1668&0.1601&4.0\\\hline
   \end{tabular}
   \end{indented}
\end{table}

\begin{figure}[t]
  \centering
  \vspace*{-2.5cm}
  \centerline{\includegraphics[width=4.0in]{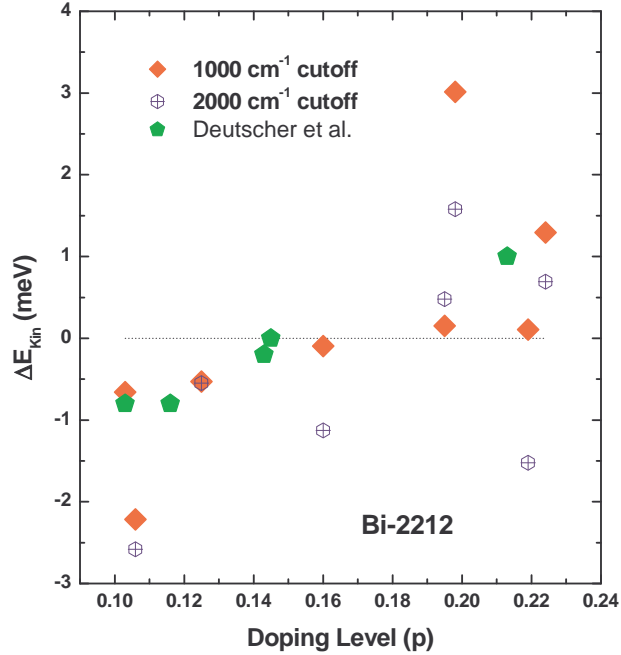}}
  \vspace*{-2.5cm}
  \caption{Doping dependent kinetic energy change,
$\Delta E_{Kin}\equiv E^s_{Kin}-E^n_{Kin}$, of Bi-2212 when the
system becomes superconductor. We also show data points from
Deutscher {\it et al.} of Bi-2212
systems~\cite{deutscher05}}\label{fig12}
\end{figure}

In figure~\ref{fig9} we display a temperature dependent $N_{eff}(T)$ of our six
representative Bi-2212 samples with a cutoff frequency of 2000 cm$^{-1}$. Above $T_c$ as
the temperature decreases $N_{eff}$ increases almost linearly and the rate of increase
grows as the doping level increases. We take into account this linear temperature dependence
of $N_{eff}$ when we adjust the superfluid density, $N_{s,sum}$ (see figure~\ref{fig12})
to obtain the kinetic energy change. We should note here that the temperature dependence
of $N_{eff}$ depends on the cutoff frequency and doping level as well. Ortolani {\it et
al.} have observed roughly $T^2$ dependence of $N_{eff}$ in La$_{2-x}$Sr$_{x}$CuO$_4$
(LSCO) systems~\cite{ortolani05}.

In figure~\ref{fig10} we show the superfluid densities obtained from the two different
methods, i.e. using Eq.~\ref{e8} and Eq.~\ref{e9}. Here we do not include the temperature
dependence of $N_{eff}$ to obtain $N_{s, sum}(\omega)$ yet. In principle, $N_{s}(\omega)$
is frequency independent and $N_{s,sum}(\omega)$ approaches a saturation value as the
frequency increases. We observe that $N_{s, sum}$ saturates more quickly as doping
increases, which is consistent with the observation of Homes {\it el al.}~\cite{homes04}
in the YBa$_2$Cu$_3$O$_{6+x}$ (Y123) system except that our saturation frequency is lower
than what they observed. The number in the parenthesis is a percentage difference between
$N_{s}$ and the saturated $N_{s,sum}$, i.e. $(N_{s}-N_{s,sum})/N_{s}\times 100$. Roughly,
we observe that the superfluid density increases as doping increases. Through the whole
spectral range and at all doping levels the saturated $N_{s,sum}$ is smaller than $N_{s}$
and the FGT sum rule is violated. However, when we take into account the temperature
dependence for $N_{s, sum}$, $N_{s,sum}$ is smaller than $N_{s}$ in underdoped region and
is larger than $N_{s}$ in overdoped (see figure~\ref{fig12}). In other words the sumrule
violation changes sign at this doping level.

In the upper panel of figure~\ref{fig11} we show a $T_{c}$ versus $\rho_{s}\equiv
\omega_{ps}^2$ graph. We do not observe so-called "boomerang"
effect~\cite{niedermayer93,villard98} where at higher doping levels the superfluid
density decreses. Also we add two additional data points for 6 \% Sr doped LSCO
($T_c$=5.5K) and a well-ordered underdoped Y123-OrthoII ($T_c$=59 K)~\cite{hwang05a}. In
the lower panel we show the doping dependent London penetration depth,
$\lambda_{L}(p)\equiv1/(2\pi \omega_{ps})$ where $\omega_{ps}$ is in \cm units. As doping
increases the London penetration depth decreases monotonically with values ranging between 1000
and 3000 \AA.
%
%

In figure~\ref{fig12} we show the doping dependent kinetic energy change, $\Delta
E_{Kin}(p)$, of Bi-2212 samples with two different cutoff frequencies; 1000 and 2000
cm$^{-1}$. When calculating $\Delta E_{Kin}(p)$ we take into account the temperature
dependent spectral weight (see Fig.~\ref{fig9}) of the normal state i.e. we extrapolate
the temperature dependent trend to get an appropriate normal state $N_{eff}$ at the
temperature of the superconducting state considered. Even though the data points do not
show a very smooth doping dependence we can clearly observe a crossover from negative
(underdoped) to positive (overdoped) kinetic energy change, which is consistent with the
result of Deutscher {\it et al.}~\cite{deutscher05} in the same Bi-2212 system. Their
data are also shown in the figure.

\section{The extended Drude model and the optical self-energy}

The extended Drude model offers a detailed description of the charge carrier scattering
spectrum and its contribution to the effective mass~\cite{puchkov96,allen71}. In this
picture the elastic scattering rate in the Drude expression is allowed to have a
frequency dependence and an extra real quantity, $\omega\lambda(\omega)$ is added to the (imaginary) scattering rate. This is necessary
to retain the Kramers-Kronig relation between $\sigma_1(\omega)$ and $\sigma_2(\omega)$.
In this formalism we can introduce an interesting and useful quantity, the optical
self-energy, $\Sigma^{op}(\omega)$, which is closely related to the quasiparticle
self-energy~\cite{hwang04,carbotte05a}.

\begin{eqnarray}
\sigma(\omega, T) &=& i \frac{\omega_p^2}{4 \pi} \frac{1}{\omega
+[\omega\lambda(\omega,T) + i 1/\tau(\omega, T)]} \nonumber\\  &=&
i \frac{\omega_p^2}{4 \pi} \frac{1}{\omega-2\Sigma^{op}(\omega,
T)}
\end{eqnarray}
where $\omega_p$ is the plasma frequency, $1/\tau(\omega,T)$ is the optical scattering
rate, and $\lambda(\omega)+1 \equiv m^*(\omega)/m$, $m^*(\omega)$ is an effective mass
and $m$ is the bare mass. The optical self-energy is a complex function,
$\Sigma^{op}(\omega, T) \equiv\Sigma^{op}_1(\omega, T) + i \Sigma^{op}_2(\omega, T)$,
where $-2\Sigma^{op}_1(\omega,T)=\omega\lambda(\omega,T)$ and $-2\Sigma^{op}_2(\omega,
T)=1/\tau(\omega, T)$. $\Sigma^{op}_1$ and $\Sigma^{op}_2$ make a Kramers Kronig pair.
The optical self-energy contains the plasma frequency, which includes the spectral weight
of the free carrier part of the optical conductivity. We obtained the plasma frequency by
using a procedure introduced in a previous study~\cite{hwang04a}. The plasma frequencies
are displayed in table~\ref{plasma}. The optical self-energy at high frequencies depends
strongly on $\epsilon_H$ (see figure~\ref{fig6} and its caption). The optical self-energy
is, apart from a $(cos\theta-1)$ factor where $\theta$ is a scattering angle, an average
over the Fermi surface of the quasiparticle
self-energy~\cite{kaminski00,hwang04,schachinger03,millis03} as measured by angle
resolved photoemission spectroscopy (ARPES). ARPES has a capability of k-space resolution
while optics has the advantage of better overall energy resolution. The self-energies
measured by the two spectroscopy techniques (optical and ARPES) show qualitatively the
same properties~\cite{hwang04,carbotte99,norman98}. However, on a quantitative level
there are some fundamental differences between them \cite{carbotte05a}.
%
%
\begin{figure}[t]
  \centering
  \vspace*{-1.5cm}
  \centerline{\includegraphics[width=4.0in]{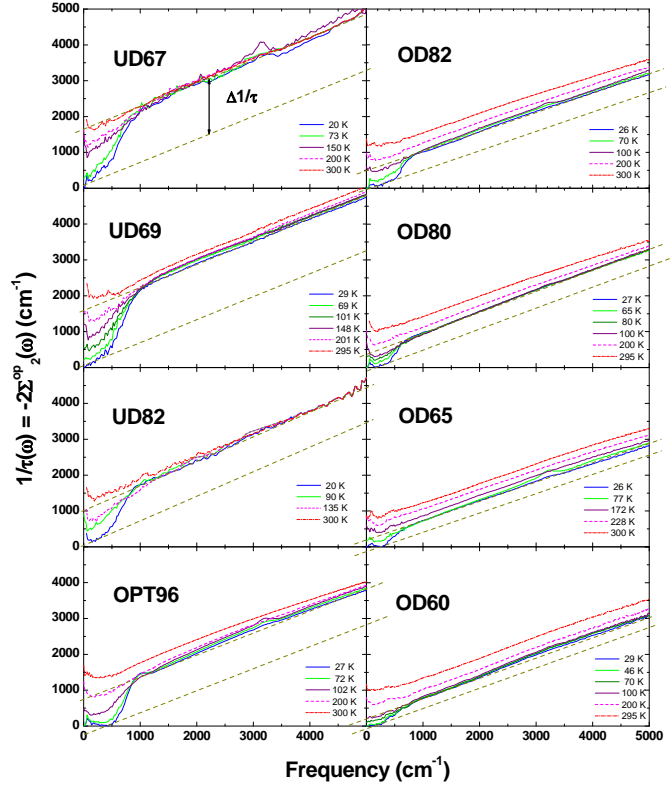}}
  \vspace*{-1.0cm}
  \caption{The optical scattering rate or the imaginary part of
the optical self-energy of ab-plane Bi-2212 at various doping
levels and temperatures. $\Delta 1/\tau$ can be a measure of
intensity of the step-like feature, which is correlated to the
magnetic resonance mode~\cite{hwang04,carbotte99,hwang05a}. The
intensity of the step shows strong temperature and doping
dependencies.} \label{fig13}
\end{figure}
%
%
\begin{figure}[t]
  \centering
  \vspace*{-2.5cm}
  \centerline{\includegraphics[width=4.0in]{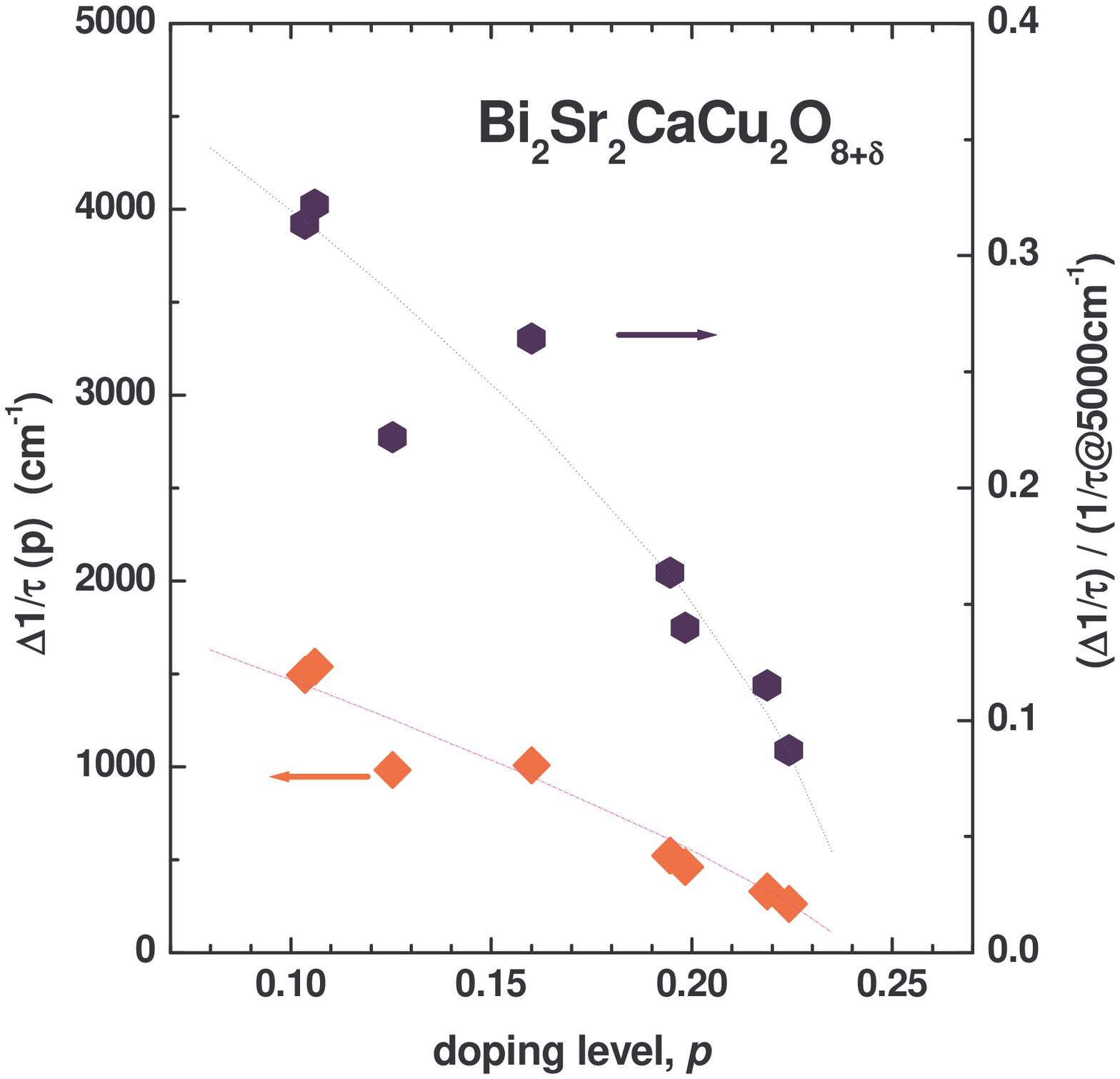}}
  \vspace*{-2.5cm}
  \caption{The doping dependent intensity of the step-like feature
in the optical scattering rate.} \label{fig14}
\end{figure}

Figure~\ref{fig13} displays the optical scattering rates of our eight Bi-2212 samples at
various temperatures above and below $T_{c}$. At room temperature the overall scattering
rate decreases as the doping increases. The scattering rate is a fairly linear as a
function of frequency and the slope decreases monotonically with the doping. This linear
dependency was described by the phenomenological marginal Fermi liquid
theory~\cite{varma89}. We do not observe any hints of "the plateau" in the scattering of
any of our samples up to 8000 cm$^{-1}$, as suggested by van der
Marel {\it et al.}~\cite{vanderMarel04}. At lower temperatures a sharp step-like feature
appears in the spectrum. The onset temperature of this feature is higher than $T_c$ for
underdoped samples and $\cong T_{c}$ for optimally and overdoped samples. This feature
can be fitted with a  formalism~\cite{hwang05a,sharapov05}, based on a method proposed by
Shulga {\it et al.}~\cite{shulga91}. For the fit we need two separate bosonic scattering
channels; one is scattering by a sharp mode and the other a broad background. Hwang {\it
et al.} showed that the step-like feature and the sub-linear part of high frequency
scattering rate were attributed to the sharp mode and the broad background in the
spectral function $\alpha^2F(\Omega)$, respectively~\cite{hwang05a}. In the
superconducting state we need a formalism that incorporates the superconducting gap and
coherence factors. Qualitatively a similar picture was derived and applied to optimally
doped Bi-2212 system by Schachinger {\it et al.}~\cite{schachinger03}. We estimate the
contribution of the sharp mode to the optical scattering rate by the height of
the step in the scattering rate, which is proportional to the area under the mode in the spectral function. We
measure the height of the step as follows~\cite{hwang04}: we draw a dashed line parallel
to the high frequency sub-linear trend from an onset point of substantial scattering. The
difference between this line and the actual high frequency scattering, $\Delta 1/\tau$
(shown in the figure~\ref{fig13}), is our estimate of the contribution of the sharp mode
to the optical scattering. A doping dependent step intensity, $\Delta 1/\tau (p)$, is
shown in figure~\ref{fig14}. In the figure we also show a normalized scattering rate,
$(\Delta /\tau)/(1/\tau@5000\mbox{cm}^{-1})$, divided by the value at 5000 cm$^{-1}$. The
step in  intensity decreases rapidly as the doping level increases and becomes zero near
doping a level of $p\sim0.24$ within the superconducting dome, where the
superconductivity is still strong in terms of $T_c\sim55$ K.

%
%
\begin{figure}[t]
  \centering
  \vspace*{-1.5cm}
  \centerline{\includegraphics[width=4.0in]{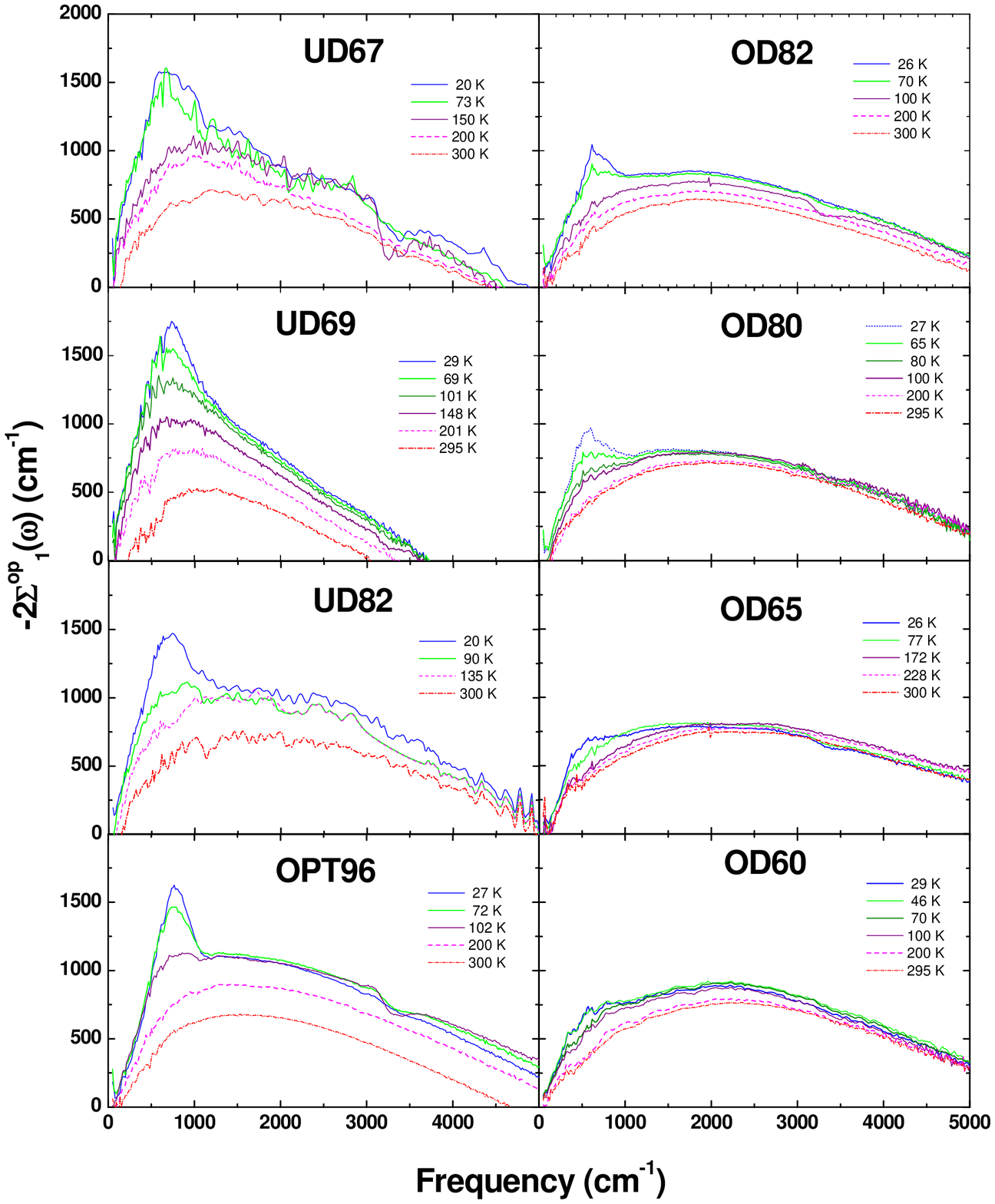}}
  \vspace*{-1.0cm}
  \caption{The real part of the optical self-energy of eight Bi-2212 samples
at various temperatures. We are able to resolve a sharp peak and a
broad background in the spectra. The sharp peak shows strong
temperature and doping dependencies.} \label{fig15}
\end{figure}
%
%
\begin{figure}[t]
  \centering
  \vspace*{-0.8cm}
  \centerline{\includegraphics[width=4.0in]{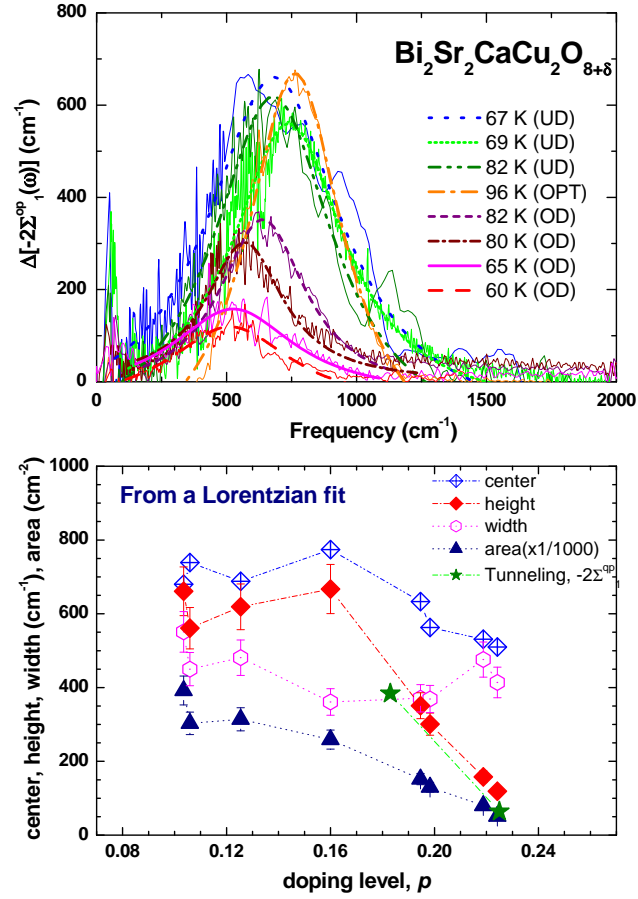}}
  \vspace*{-0.6cm}
  \caption{In the upper panel we show the sharp mode separated from
the high temperature broad background in the real part of the
self-energy and a Lorentzian fit for each doping level. In the
lower panel we show the center frequency, the height, the width of
the sharp mode and the area under the mode obtained from the
Lorentzian fit. The two star symbols are extracted from the
self-energy in a recent tunnelling study by Zasadzinski {\it el
al.}~\cite{zasadzinski06,hwang06}.} \label{fig16}
\end{figure}

%
%
\begin{figure}[t]
  \centering
  \vspace*{-0.8cm}
  \centerline{\includegraphics[width=4.0in]{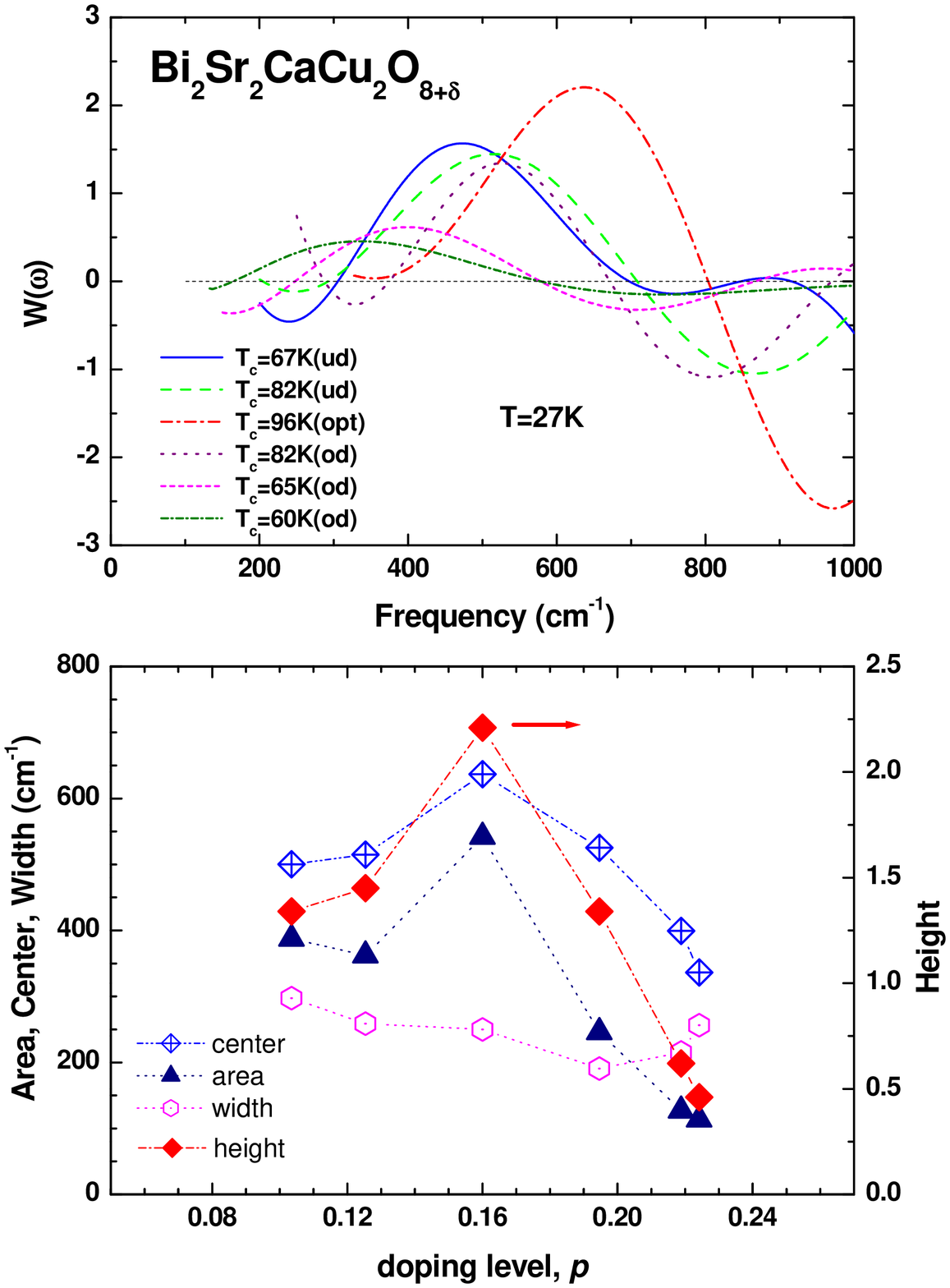}}
  \vspace*{-0.6cm}
  \caption{In the upper panel we show the bosonic spectral function,
$W(\omega)$, obtained by using the second derivative formalism
(see Eq. 17) at a superconducting state, $T\sim$ 27 K. We observe
sharp mode at low frequency range. In the lower panel we show the
center frequency, the height, the width of the sharp mode and the
area under the mode in $W(\omega)$.} \label{fig17}
\end{figure}

In figure~\ref{fig15} we display the real part of the optical self-energy of Bi-2212.
This quantity and the scattering rate make a KK pair. At room temperature we see only a
very broad background spectrum. The peak frequency of the broad background increases as
the doping increases. As the  temperature is lowered a well-defined sharp peak appears
out of the broad room temperature background with the same onset temperature as the
step-like feature in the scattering rate and the peak grows as the temperature is reduced
further. Johnson {\it et al.} have also resolved a sharp peak and a broad background in
ARPES quasiparticle self-energy spectra in the ($\pi$,$\pi$) direction of Bi-2212 systems
and correlated the sharp mode to the magnetic resonance mode of inelastic neutron
scattering~\cite{johnson01}. More recently similar conclusions have been reached by
Kordyuk {\it et al.}~\cite{kordyuk05}. The optical self-energy and ARPES quasiparticle
self-energy show strong similarity in terms of their shapes and temperature and doping
dependencies~\cite{hwang04}. Both self-energies have a sharp mode and a broad background
at low temperature below $T_c$ while optics shows much better energy resolution. The mode
frequencies from both techniques have an interesting relation:
$\Omega_{optical}\cong\sqrt{2}\:\:\:\Omega_{ARPES}$, where $\Omega_{optical}$ and
$\Omega_{ARPES}$ are the optical and ARPES mode frequencies, respectively. The relation
is consistent with a theoretical prediction~\cite{carbotte05a}. Furthermore Hwang {\it et
al.} have analyzed temperature dependent intensity of the step-like feature in the
optical scattering rate of Y123-OrthoII and found a direct correlation between the step
feature in the scattering and the magnetic resonance mode of inelastic neutron scattering
(INS)~\cite{hwang05a}.

In the upper panel of figure~\ref{fig16} we show the sharp peak separated from the broad
background in the real part of the optical self-energy and fitted it to a Lorentzian
function to obtain the center frequency, the width, and the height of the peak and the
area under it. For the two most underdoped samples it is a little difficult to separate
the peak from the broad background because the peak evolves gradually from the broad peak
of the background. We display the doping dependent properties of the peak in the lower
panel of the figure. The center frequency (the width) seems to be maximized (minimized)
near the optimally doping level. Note that the width does not change much through a wide
range of doping. The area and height show a strong doping dependence in the overdoped
region; the intensities of both quantities decrease very rapidly as doping increases and
finally become zero simultaneously within the superconducting dome with a hole doping
level, $p \sim 0.24$ estimated from extrapolations. This result is consistent with the
doping dependent intensity of the step-like feature in the scattering rate. In the same
panel we show two data points (stars) extracted from the quasiparticle self-energy in a
recent tunnelling study by Zasadzinski {\it et al.}~\cite{zasadzinski06,hwang06}.

For another measure of the strength of the interaction of the charge carriers with the
bosonic sharp mode we use the procedure introduced by Marsiglio {\it et al.}, Carbotte
{\it et al.}, and Abanov {\it et al.}~\cite{carbotte99,marsiglio98,abanov01} where the
bosonic spectral function, $W(\omega)$, is derived from the second derivative of the
optical scattering rate times the frequency. This function can be described as
follows~\cite{tu02,allen71,marsiglio98}:
\begin{equation}
W(\omega)\equiv\frac{1}{2\pi}\frac{d^2}{d\omega^{2}}
\bigg[\frac{\omega}{\tau(\omega)}\bigg]
\end{equation}
and $W(\omega)\approx \alpha^2 F(\omega)$ at zero temperature in the normal state, where
$\alpha$ is a coupling constant, and $F(\omega)$ is a bosonic density of states. To
obtain the spectral function ($W(\omega)$) we followed a smoothing procedure introduced
by Tu {\it et al.}~\cite{tu02}. We fit $1/\tau(\omega)$ (see figure~\ref{fig13}) with a
polynomial with ten terms to catch the main frequency in the spectra without including
too much experimental random-noise.

%
%
\begin{figure}[t]
  \centering
  \vspace*{-0.8cm}
  \centerline{\includegraphics[width=3.5in]{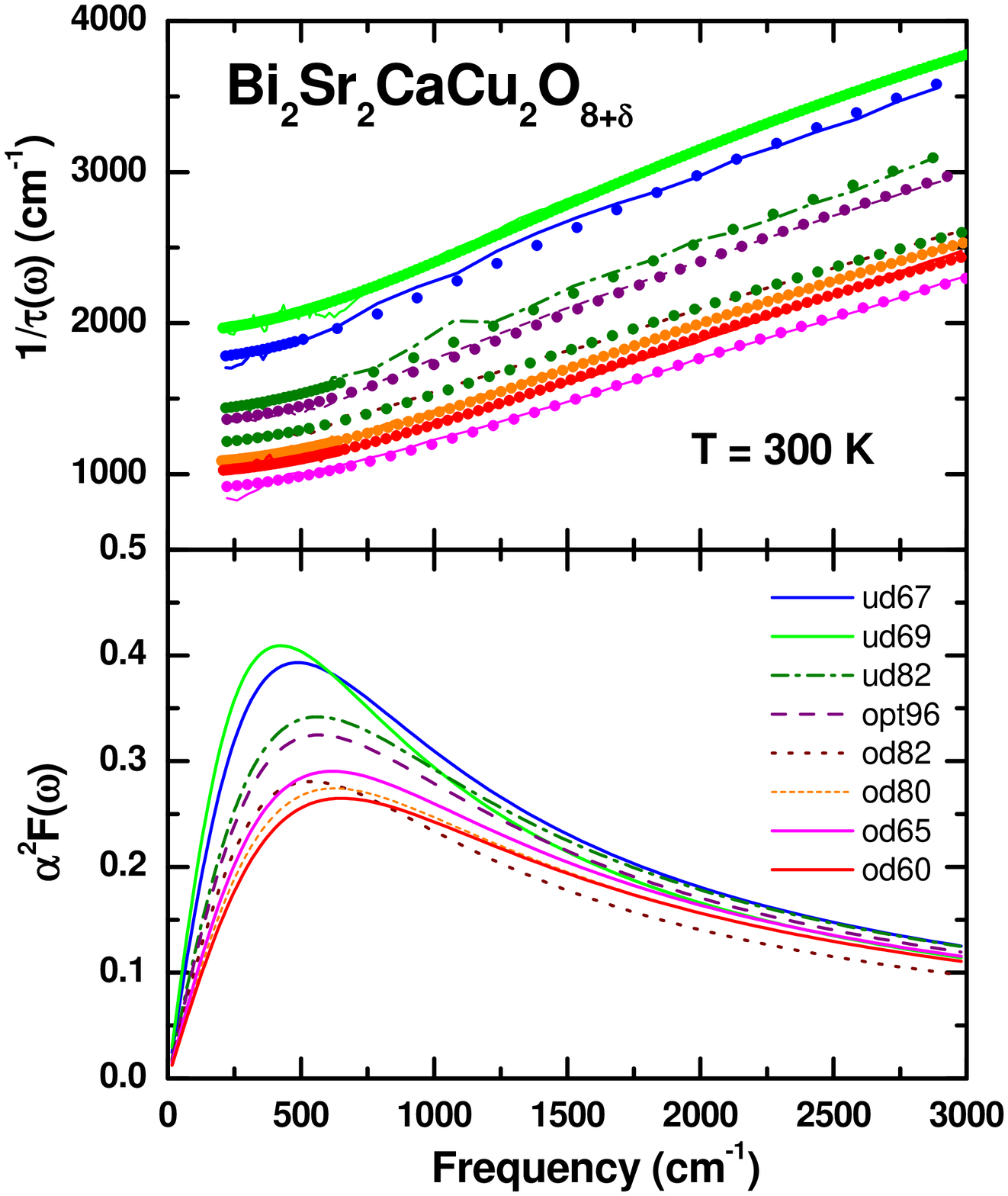}}
  \vspace*{-1.6cm}
  \caption{In the upper panel we show that the optical scattering
rates (lines) of our eight Bi-2212 samples at room temperature and their fits (symbols)
by using the methods introduced in a literatures~\cite{hwang05a,sharapov05}. In the lower
panel we display the MMP bosonic spectral function, $\alpha^2F(\omega)$, from the fits in
the upper panel.} \label{fig18}
\end{figure}
%
%
\begin{figure}[t]
  \centering
  \vspace*{-2.5cm}
  \centerline{\includegraphics[width=3.5in]{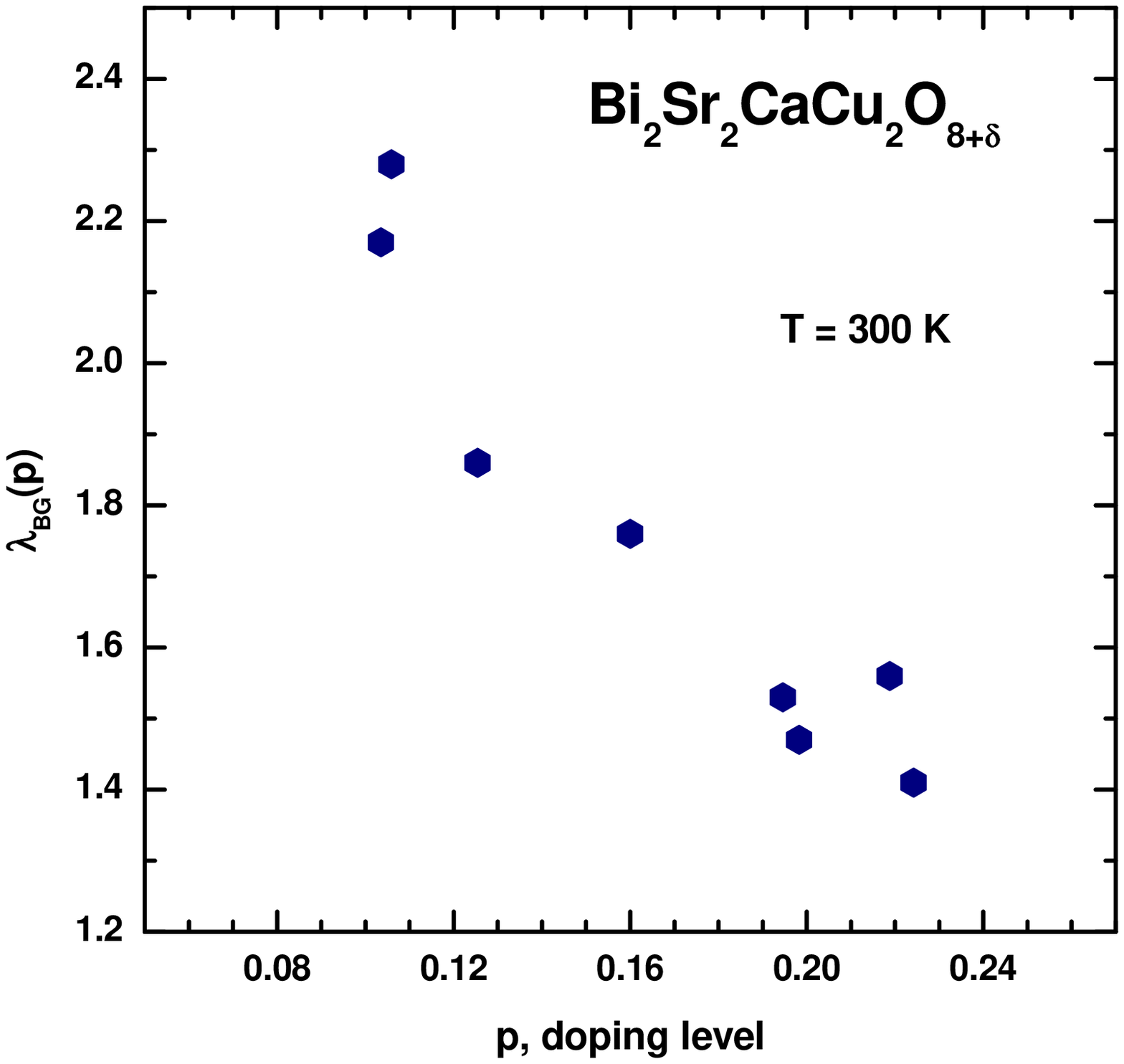}}
  \vspace*{-2.5cm}
  \caption{Doping dependent mass enhancement factor, $\lambda(p)
=2\int^{\omega_c}_{0}\alpha^2 F(\omega')/\omega'\:d\omega'$, where $\omega_c$=3000
cm$^{-1}$, at room temperature. We used the extracted spectral function,
$\alpha^2F(\omega)$, in the lower panel of figure~\ref{fig18} to get the coupling
constant, $\lambda(p)$.} \label{fig19}
\end{figure}

In the upper panel of figure~\ref{fig17} we show the bosonic spectral function obtained
using the second derivative method, $W(\omega)$ for our six representative Bi-2212
samples at their lowest temperature. In the figure we have subtracted the room
temperature background $W(\omega)$  from $W(\omega)$ at the lowest temperature. We should
note that the bosonic mode is enhanced in the superconducting state. However, the
qualitative doping dependent trend is not affected by the gap~\cite{carbotte05a}. In the
lower panel of the figure we display the doping dependent  properties of the peak: the
center frequency, the width, and the height of the peak, and the area under the peak. The
area is roughly proportional to the quantity, $\Delta 1/\tau$; it roughly follows the
same doping dependent trend as $\Delta 1/\tau$. The extrapolation of the variation with
doping of the height and area suggests that the mode will disappear within the
superconducting dome in the Bi-2212 phase diagram. The polynomial fit method is fairly
robust with respect to experimental noise.  We compared this method with alternate
methods including an inverse matrix method~\cite{dordevic05} and found that in the Ortho
II material at least, this method gave a well ordered temperature sequence for both the
peak position and frequency where alternate methods tended to scramble these quantities.

Since the mode disappears above a critical doping level in the phase diagram leaving the
broad background as the only component in the interacting bosonic spectrum it is
important to investigate this spectrum. We obtained the bosonic spectral functions of our
eight Bi-2212 samples by fitting the room temperature scattering rate with the
Millis-Monien-Pines (MMP) type background~\cite{millis90} using the methods introduced by
Sharapov {\it et al.}~\cite{hwang05a,sharapov05}. In the upper panel of
figure~\ref{fig18} we show the fits (symbols) and data (lines) of the optical scattering
rate. In the lower panel of the figure we display the corresponding  bosonic spectral
functions. The spectral function shows a slight doping dependence at low frequencies. As we
move closer to the antiferromagnentic phase boundary the spectral function grows in
strength and develops a more defined peak at low frequency. We also obtain the coupling
constant, $\lambda$, for this spectrum, which is defined by
$\lambda=\int_{0}^{\omega_c}\alpha^2F({\omega'})/\omega'\:\:d\omega'$ with $\omega_c$ =
3000 cm$^{-1}$. The resulting doping dependent coupling constant, $\lambda_{BG}(p)$, is
shown in Fig.~\ref{fig19}. There is a strong doping dependence in the coupling constant,
range from 2.3 to 1.4. The value decreases monotonically as doping level increases. This
behavior is not consistent with an observation of Padilla {\it et al.} in LSCO
systems~\cite{padilla05} who suggest that there is little doping dependence of the
carrier mass which is given by $m^*=m(1+\lambda)$ across the phase diagram.

\section{Discussion and conclusions}

\subsection{The superfluid density and the kinetic energy change}

The doping dependence of the superfluid density in the cuprates has been of great
interest from early on when Uemura {\it et al.} found using muon scattering that the
superfluid density increased with doping and was proportional to the superconducting
transition temperature~\cite{uemura91}. This so called Uemura relationship holds only up
to the optimal doping level at which point as $T_c$ stops increasing while the superfluid
density continues to increase. There have been reports for example by Niedermayer {\it et
al.} who studied overdoped Tl-based (Tl$_2$Ba$_2$CuO$_{6+\delta}$: Tl-2201) cuprates by
muon spin resonance of a so-called boomerang effect where the superfluid density {\it
decreases} in the overdoped region as the doping level {\it
increases}~\cite{niedermayer93}.  As we see in Fig.~\ref{fig11} our data, obtained by two
different methods, shows that the superfluid density continues to increase in the
overdoped region and there is no boomerang effect in this material.


In conventional superconductors the Ferrel-Glover-Tinkham (FGT) sum rule holds. However,
the c-axis optical transport in the cuprates shows a strong violation of the FGT sum rule
which has been attributed to a kinetic energy driven superconductivity~\cite{basov99}.
Whether the FGT sum rule holds or not in ab-optical transport has been the subject of
some controversy~\cite{boris04,kuzmenko05,santander04}. Here we find, as shown in
Fig.~\ref{fig12}, that a measurable FGT sum rule violation can be observed in ab-plane
transport of Bi-2212. We do find that we have to take account the temperature dependence
of the spectral weight and extrapolate it into the superconducting state. This result is
consistent with Deutscher {\it et al.}~\cite{deutscher05}.

\subsection{The sharp mode and the broad background in the optical self-energy}

We have verified the behavior of the sharp mode observed by Hwang
{\it et al.}~\cite{hwang04} over a wide range of dopings with more
samples which is shown in Fig.~\ref{fig14} and Fig.~\ref{fig16}
and with a different method of analysis, the second derivative
method, the result of which is shown in Fig.~\ref{fig17}.

We also have studied the broad background in more detail; we
extracted doping dependent bosonic spectra shown in
Fig.~\ref{fig18} from room temperature optical scattering rates.
The broad peak of the background moves to lower frequencies and
its intensity increases as doping decreases, which is
qualitatively consistent with Fig. 51 of Ref.~\cite{eschrig06}.
The coupling constant of the room temperature background,
$\lambda_{BG}(p)$, which is displayed in Fig.~\ref{fig19},
decreases significantly as doping increases which agrees with
previous studies~\cite{johnson01,hwang04a}.

\subsection{Summary and conclusion}

We obtained the superfluid density with two different methods from
our ab-plane optical conductivity and observed that the superfluid
density increased monotonically with hole doping level. The smooth
and monotonic increase of superfluid density with doping supports
an overall consistency of our study. We took into account the
temperature dependence in the optical spectral weight to
extracting the superfluid density and observed a violation of the
FGT sumrule, the result of which causes a change of the kinetic energy
of the charge carriers at the superconducting transiotion. Kinetic
energy increases (decreases) in overdoped (underdoped) systems as
the system becomes a superconductor. We also confirmed our
previous work (Ref.~\cite{hwang04}) with more samples and a
different way of analysis. We resolved a sharp mode out of a broad
background in the optical self-energy. The temperature and doping
dependence of the sharp mode is dramatic; the onset temperature
$T_s$ of the sharp mode is above (at) $T_c$ in underdoped
(overdoped) systems and the intensity of the mode gets weakened
strongly in the overdoped region with increasing doping and an
extrapolation of the doping trend shows a complete disappearance
above a critical doping, $p_c\sim$ 0.24, within the
superconducting dome. The broad background presents at all
temperatures and doping levels and shows relatively weak doping
and temperature dependence. However the coupling constant of the
background decreases measurably as doping increases.

\ack

This work has been supported by the Canadian Natural Science and Engineering Research
Council and the Canadian Institute of Advanced Research. We thank H. Eisaki and M. Greven
for supplying us with several crystals. Their work at Stanford University was supported
by the Department of Energy's Office of Basic Sciences, Division of Materials Science.
The work at Brookhaven was supported in part by the Department of Energy.

\end{document}